\documentclass[12pt,preprint]{aastex}
\newcommand{\etal}{{et al.}}
\newcommand{\ie}{{i.e.,}}
\newcommand{\eg}{{e.g.,}}

\newcommand{\ergflx}{erg cm$^{-2}$ s$^{-1}$}

\begin{document}

\title{The Aligned $z \sim 1$ Radio Galaxy 3C\,280\footnotemark[1]}

\author{Susan E. Ridgway\altaffilmark{2}}
\affil{Johns Hopkins University, Dept. of Physics and Astronomy\\
3400 N. Charles St. Baltimore, MD 21218}
\email{ridgway@pha.jhu.edu}
\author{Alan Stockton\altaffilmark{2,3}}
\affil{Institute for Astronomy, University of Hawaii, 2680 Woodlawn Drive,
Honolulu, Hawaii  96822}
\email{stockton@ifa.hawaii.edu}
\author{Mark Lacy}
\affil{SIRTF Science Center, MS 220-6, Caltech, 1200 E. California Blvd.,
Pasadena, CA 91125}
\email{mlacy@ipac.caltech.edu}

\footnotetext[1]{Based on observations made with the NASA/ESA Hubble Space 
Telescope, obtained at the Space Telescope Science Institute, which is 
operated by the Association of Universities for Research in Astronomy, Inc., 
under NASA contract NAS 5-26555.}

\altaffiltext{2}{Visiting Astronomer, W.M. Keck Observatory, jointly operated
by the California Institute of Technology and the University of California.}

\altaffiltext{3}{Visiting Fellow, Research School of Astronomy and Astrophysics,
The Australian National University, Private Bag, Weston Creek P.O., Canberra,
ACT 2611, Australia}

\begin{abstract}
\tighten

The $z\sim1$ radio galaxy 3C\,280
has a particularly striking rest-frame UV morphology, with 
multiple line and continuum components precisely aligned with the
radio structure, including an obvious semi-circular arc.
Here we explore the nature of these various
components by bringing together {\it HST} and ground-based imaging, ground-based
spectroscopy, and radio mapping. 
From plausible decompositions of the spectra,
we show that the continuum of the nuclear component is likely dominated by
a combination of nebular thermal continuum, quasar light, and light from
old stars.  A component that falls directly on the probable
path of the radio jet shows mostly nebular thermal
continuum and includes contributions from 
a relatively young stellar population with age around 100 Myr.
The arc appears to be completely dominated by line emission and
nebular thermal continuum, with no evidence for a significant stellar
contribution. 
Though much of the aligned light is in UV components, 
the underlying old elliptical is also well-aligned with the radio axis. 
The elliptical is well-fit by a de Vaucouleurs profile, probably 
has a moderately old stellar population ($\sim$ 3 Gyr), and is
a massive system with a velocity dispersion of $\sigma \approx$ 270 km s$^{-1}$
that implies it contains a supermassive black hole. 
Although the arc and the extended emission 
surrounding the eastern lobe suggest that interactions
between the radio lobe and jet must have been important in
creating the UV morphology, 
the ionization and kinematic properties in these components
are more consistent with photoionization than shock excitation.
3C\,280 may be a transition object between the compact steep-spectrum
radio galaxies which seem to be shock-dominated, and the extended 
radio sources which may have evolved past this phase
and rarely show shock signatures. 

\end{abstract}

\section{Introduction}

Our understanding of the physical processes underlying the
detailed morphologies of high-redshift radio galaxies, including
the well-known radio-optical alignment effect
\citep{cha87,mcc87}, remains unclear.
In the near-infrared, most $z \sim 1$ powerful radio galaxies appear to be
relaxed, giant elliptical galaxies 
(\eg\ \citealt{rig92,bes96}), and high-resolution $H$ band imaging with 
the {\em HST} NICMOS camera confirms this result \citep{zir99, zir03}.
In the optical (rest-frame UV), 
a variety of types of alignment have been found in
WFPC2 imaging of $z\sim1$ radio galaxies
\citep{dic95,lon95,bes96,bes98,rid97}.
Aligned rest-frame UV continuum structures that at ground-based resolution
appear smooth or to consist of multiple large-scale components
have been resolved by WFPC2 imaging
in many cases into a sequence of discrete, almost unresolved
peaks, closely confined to the radio axis
(\eg\ 3C 324 and 3C 368; 
\citealt{dic95,lon95}).
In 3C 368, the continuum is fairly blue and unpolarized
\citep{vBr96},
and many of the components are dominated by thermal emission from the
emission-line producing gas
\citep{dic95,sto96}.
In 3C 324, however, the chain of lumps exhibits a mild curvature,
presumably associated with the precession of the radio jet.
Detection of polarization in this and many other high-$z$ radio galaxies 
supports the hypothesis that the bulk of the UV radiation comes
from scattered light (\eg\ \citealt{cim96,ver01}). High resolution
HST polarization studies of 3C 324 have shown that each well-aligned
subclump is in fact highly polarized ($\gtrsim$12\%), consistent
with dust-scattered quasar light contributing 10--40\% in
the rest-frame UV, and even more in the rest-frame optical \citep{zir03t}. 
However, the mechanisms that might
confine the scatterers to the jet path are not well understood.
In these cases, the morphologies argue 
for the interaction of the jet with the ambient medium as the primary
cause of the structures observed. A correlation seen between
the size of the radio structure and the tightness of the
alignment can be seen as evidence in favor of jet-induced star formation
\citep{bes96}, and direct evidence for young stellar features
has been found in one high-$z$ radio galaxy \citep{dey97}. 
More than one of these mechanisms
may be important in any particular aligned radio galaxy, and it is
probably necessary to understand many high-$z$ radio galaxies in terms of
a combination of these processes.

We have obtained some particularly intriguing examples of radio-optical
alignment as a result of a WFPC2 program
to image a complete sample of $z\sim1$ 3C radio sources
\citep[henceforth RS97]{rid97}.
One of these, the $z=0.998$
radio galaxy 3C\,280, has an aligned morphology that is unique
among {\em HST} images of high-$z$ radio galaxies and seems quite difficult
to explain with any combination of the commonly considered mechanisms
for optically aligned continuum emission.
Its arc-like rest-frame UV morphology has been noted and
discussed in \citet{bes96,bes98} and in RS97. Here we present the
results of a program to use multicolor
imaging, optical and infrared spectroscopy,
and MERLIN and VLA radio maps to make a
detailed study of this unusual powerful radio galaxy.
We will use $H_0=70$
km s$^{-1}$ Mpc$^{-1}$, $\Omega_m = 0.3$, and
$\Omega_{\lambda}=0.7$ throughout.

\section{Observations and Reduction}
\subsection{Imaging data}
We give in Table \ref{obslog} a summary of the imaging data we
will use in this paper.
From RS97 we use the deep WFPC2 F622W observations of 3C\,280, as well
as the deep ground-based $K'$ and [\ion{O}{2}] images from
Keck and CFHT respectively.
The observations and reduction of these data are discussed in RS97.
From the {\em HST} archive we have also obtained 
F814W WFPC2 observations and F160W NICMOS observations.
To reduce the 
WFPC2 data, we took the best recalibrated pipeline images and, 
as these observations
consisted of a few individual undithered exposures, 
we used the STSDAS package {\it crrej} directly
to remove the cosmic rays and combine the data. 
The F814W image was taken with the WFC detector with a pixel
scale of 0\farcs1 pixel$^{-1}$.

The F160W NICMOS data from the {\em HST} archive consisted of a standard
dithered sequence of 8 exposures. We calibrated the raw data 
using the recommended bias, dark and flat fields, and also
removed the bias ``pedestal'' using software provided by
M. Dickinson. To align the separate exposures, we used offsets from
the instrument jitter files, masked out cosmic rays, and combined
the images after resampling by a factor of two resulting
in a pixel scale of 0\farcs0375 pixel$^{-1}$.
The FWHM of a star in the
final frame is 0\farcs14, yielding an image resolution very similar
to that in the WFPC2 images. 

\subsection{Optical spectroscopy}\label{optspec}

\begin{figure}[!bt]
\epsscale{0.9}
\plotone{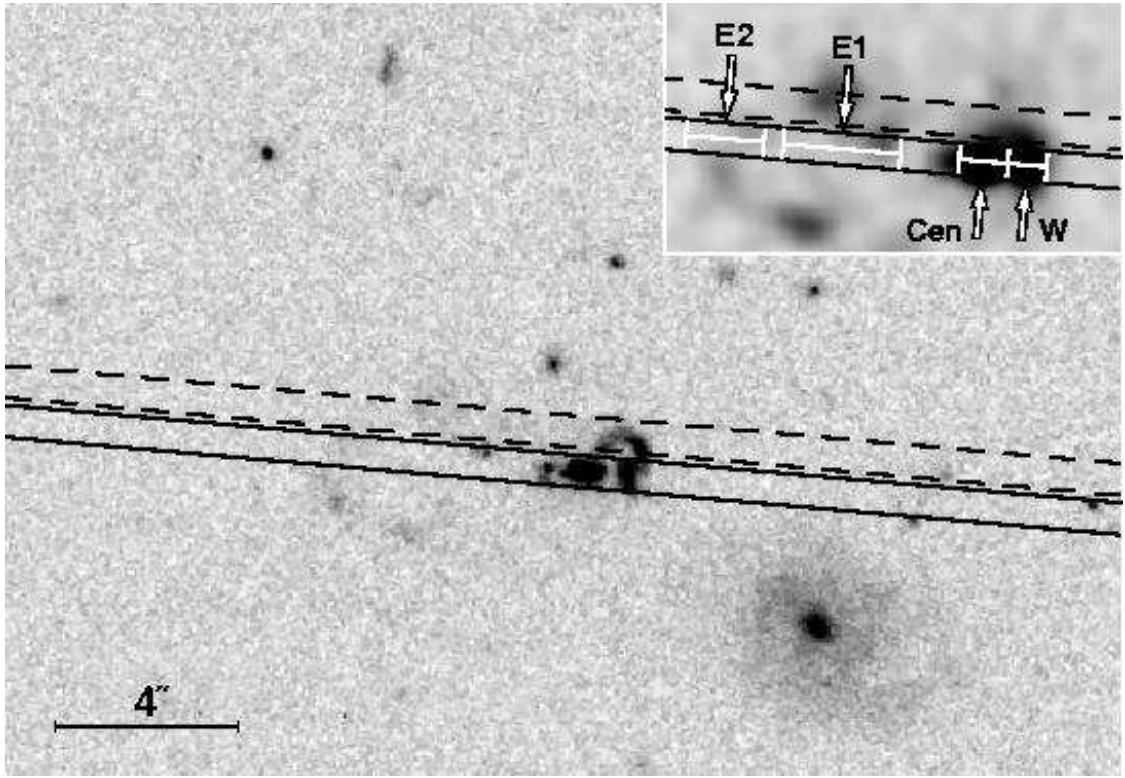}
\figcaption{{\it HST} WFPC2 image of 3C\,280 with 0\farcs7 slits used
for the Keck LRIS observations superposed (some observations
were taken in 1\arcsec\ slits).  The inset shows the [\ion{O}{2}]
image obtained with the CFHT. 
Some of the features included in spectral extractions 
shown in subsequent figures are labelled; the white bars show
the spatial aperture used to extract each spectrum.
\label{slitfig}}
\end{figure}

We obtained optical spectroscopy of 3C\,280 on the nights of
1996 February 14 and 15 with the Low Resolution Imaging Spectrometer
\citep{oke95} on the Keck I telescope. The detector was
a Tek $2048\times2048$ CCD with 0\farcs215 pixel$^{-1}$.
We used a position angle of 85\arcdeg\ for all observations,
corresponding to the axis of the primary galaxy components $a$, $b$ and $c$
(labelled in Fig.~\ref{images}).
On 14 Feb the atmospheric seeing conditions were consistently very good,
and we used the 0\farcs7 width slit throughout.
For these observations,
we used a 600 groove mm$^{-1}$ grating blazed at 7500 \AA,
covering the wavelength range 5945--8506 \AA.
We took two 1200 s spectra with the slit centered on the radio galaxy nucleus,
and two 1200 s exposures with the slit offset 0\farcs9 north, parallel to
the first slit position, in order to emphasize the contribution from
the arc $d$.
These slit positions and widths are shown in Fig.~\ref{slitfig}.


On 15 Feb, we obtained three 1200s exposures centered at the nucleus through
the 0\farcs7 width slit and with the same red
grating (600 groove mm$^{-1}$, 7500 \AA\ blaze) as the first night,
but slightly offset in wavelength coverage (6645--9235 \AA).
In addition, we obtained three 1200 s exposures centered at the nucleus
using a blue grating (600 grooves mm$^{-1}$, blaze at 6000 \AA), covering
a wavelength region of 4325--6890 \AA, and beginning with the
same 0\farcs7 slit width.
The last two of these exposures were taken with a slightly wider slit width
of 1\farcs0 because of degenerating seeing conditions.
We also obtained with this setup a spectrum of a nearby field star
to aid in deconvolution. When obtaining multiple exposures at a single position
we dithered slightly along the slit in order to aid in removal of chip defects
and cosmic rays.

The spectra were reduced using standard IRAF tasks. After initial processing
of the CCD frames, the spectra were wavelength calibrated using
observations of arc lamps and night sky lines, then were corrected
for spectral and spatial distortions, linearizing the wavelength scale.
Background sky was removed by fitting cubic splines
to the night sky lines and subtracting.
The spectra were flux calibrated and corrected for atmospheric
absorption using the spectrophotometric standards G191B2B and HZ 44.
All the spectra for the slits centered on the nucleus
were combined to make a single two-dimensional
spectrum covering the wavelength region 4325--9235 \AA; this
required scaling the spectra to
match the continuum values in regions of spectral overlap before averaging.
Average continuum values in the spectra taken
with the narrower slits were scaled up by $\sim$40--50\% to match those
in the wider slit spectra, for example,
but there were no continuum
slope discontinuities in the overlap regions and no changes in slope were made.
We also use a direct average when we combine spectra taken with 
the two different slit widths, giving the combined spectrum in those
spectral regions an average of the two spectral resolutions. 

A slight residual spatial-spectral curvature was removed from the
full-length 2-D spectrum by extracting 1-pixel-wide apertures
parallel to the trace of the combined continuum.
The spectral pixel scale is 1.26 \AA\ pixel$^{-1}$, and the spectral
resolution (as measured from arc lines) is $\sim$ 3.5 \AA\ for the 0\farcs7
slit spectra and $\sim$ 4.0 \AA\ for the 1\farcs0 slit width.
Wavelength errors are $\sim$ 1 \AA, and statistical errors in the fluxes
are $\sim$ 10\%.
The spatial resolution was measured from the continua of the calibration
stars observed near in time to the galaxy spectra, and averaged
0\farcs86.

\subsection{Near-Infrared Spectroscopy}

Near-infrared spectroscopy of 3C\,280 was performed with CGS4 on UKIRT. $J$-band
observations centered on 1.325$\mu$m were taken on 1998 January 22 UT with a
1.2\arcsec\ (two pixel) wide slit at PA 85\arcdeg. The 
150 line/mm grating was used in 3rd order, giving a resolving power of 2350. 
The total integration time was 120 min, 
split into pairs of 5 minute exposures dithered along the slit in the 
standard ABBAABBAA... sequence (Eales \& Rawlings 1993). Further 
CGS4 observations, centered at 0.98$\mu$m to include the 
[\ion{O}{3}]~$\lambda5007$
line, were taken on 1999 May 15 UT. 
The 150 line/mm grating was used in 4th order,
giving a resolving power of 3100. Total integration time was 100 minutes, with 
the data being taken in the same way as the 1998 data.

\subsection{MERLIN and VLA Radio Maps}

Though we have obtained multifrequency continuum polarization 
maps at both MERLIN and the VLA of this object, here we will discuss
only the MERLIN L band (1.4 GHz) map (to provide the best match in resolution
to the {\em HST} images) and a VLA X band (8.4 GHz) map that we obtained from the 
VLA archive. The rest of the radio mapping and depolarization 
studies will be presented in a future paper. 
The MERLIN observations consisted of 
one 12 hour run from 12 April 1997 and were self-calibrated and mapped 
in the standard way
using AIPS and {\it difmap}. The map was reconvolved
with a slightly elliptical beamsize of 0\farcs22 $\times$ 0\farcs19 FWHM. 
In this map there was no core detection. 

The VLA X band observation was taken from the VLA archive, program
AO105.
This 8.4 GHz observation was taken at A array, 
and consisted of a 6 minute snapshot. The data were phase and
flux calibrated with AIPS, then self-calibrated using {\it difmap}.
After several iterations of self-calibration, we were able to
detect the core, which exhibits some extension in the lobe 
direction that may correspond to a portion of the jet.
The map has been restored with a circular 0\farcs4 beam.
The core center is at
RA = $12^{\rm h} 56^{\rm m} 57\fs699$, Dec = 
+47\arcdeg20\arcmin19\farcs92 (J2000) and we use this
location to align the MERLIN radio map to the optical and infrared images.

\section{Analysis and Results}

\subsection{Radio mapping}

\begin{figure}[!bt]
\epsscale{0.9}
\plotone{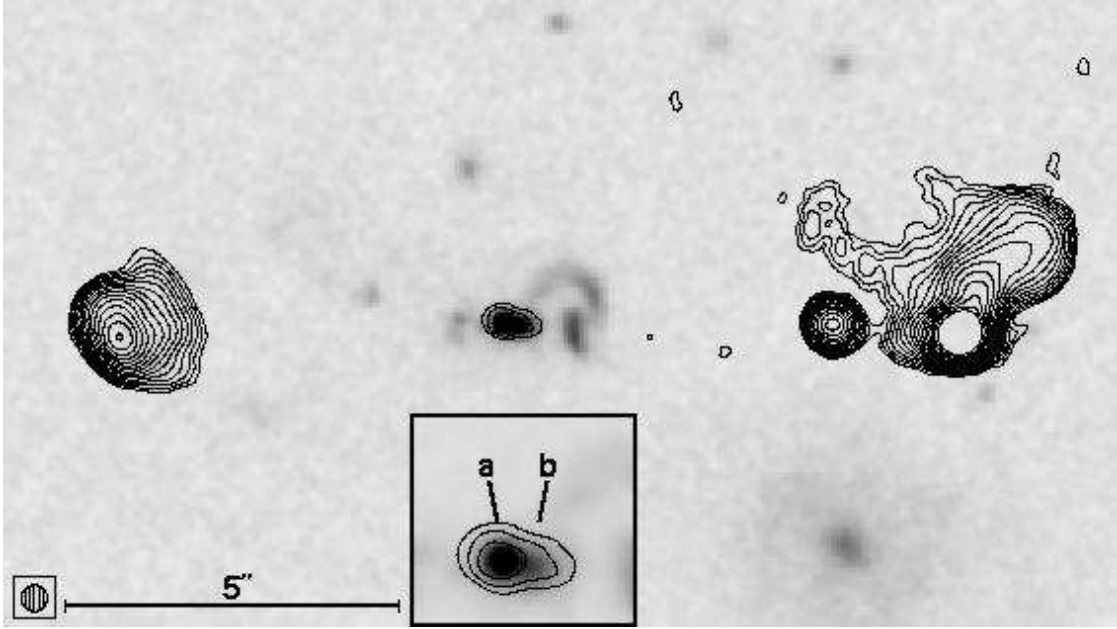}
\figcaption{VLA 8.4 GHz contour map showing the resolved radio
core, superimposed on the F622W image of 3C\,280. 
The lowest contour level is 3.5 mJy beam$^{-1}$, and the levels increase
by factors of $\sqrt{2}$.
The 0\farcs4 round beam is shown in the lower left corner.
The inset shows the core region of the F622W image and X-band map 
of 3C280 at two times the scale
of the larger image; the box is 1\farcs6 $\times$ 1\farcs7.
We have labelled the optical core components $a$ and $b$ as also labelled
in Fig.~\ref{images}. 
\label{xbandmap}}
\end{figure}

\begin{figure}[!bt]
\epsscale{0.9}
\plotone{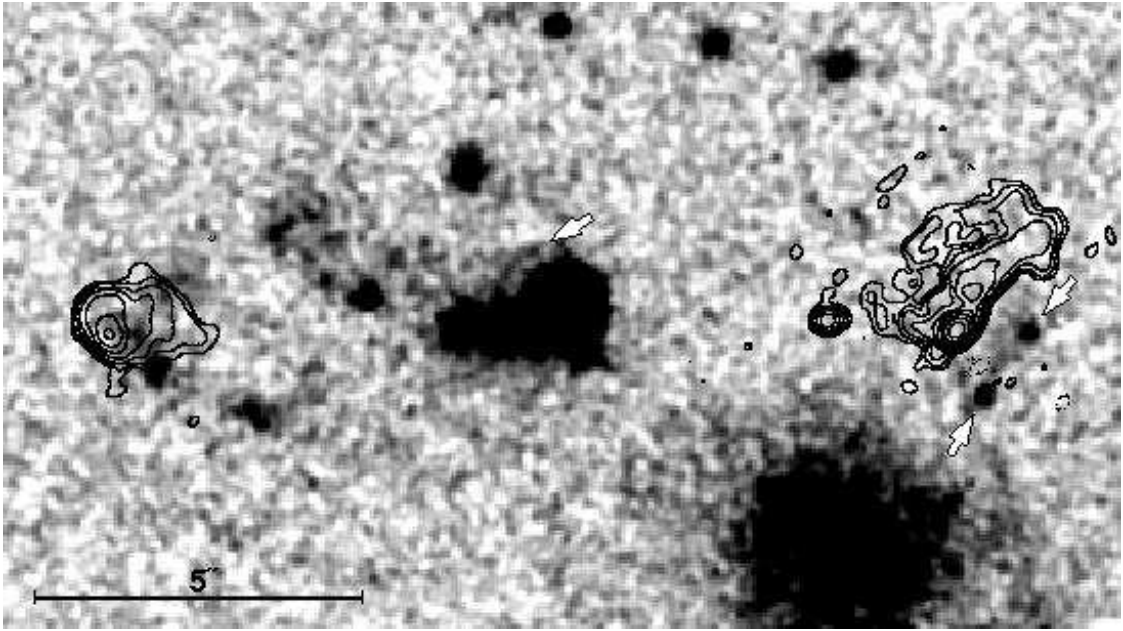}
\figcaption{MERLIN 1.4 GHz contours superposed on the {\it HST} WFPC F622W
image. No radio core was detected in this radio map. 
The lowest positive contour level (solid line) is 2.5 mJy beam$^{-1}$, and 
the restoring beam size is 0\farcs22 $\times$ 0\farcs19.
The {\it HST } image is shown at a deeper stretch to allow
lower surface brightness features to be seen. 
A secondary arc feature in the optical image is indicated with a white
arrow.
\label{merlincont}}
\end{figure}

In Fig.~\ref{xbandmap},
we show the contours of the X band (8.4 GHz) VLA map over 
the WFPC2 F622W image, and in
Fig.~\ref{merlincont} we show the L band (1.4 GHz) MERLIN map
over the same F622W image at a different intensity scaling. 
The core
is clearly detected in the X band map, and this position allows
us to align the radio and optical frames quite precisely. 
The X band core is resolved, and it is elongated to the west 
at a PA of about 85\arcdeg\ in the direction of the western hotspots and
lobe. 
The elongation of the core to the west may be emission
from the base of the radio jet, and the compact knot and
hotspot seen on the western side suggest that the radio
jet emission on this side is Doppler boosted. 
The western side is also 
less depolarized than the eastern one \citep{liu91}.
All of these indications are consistent with the western lobe being closer
to the observer than the eastern. 


The rest-frame UV morphology of the central galaxy is very well aligned
with this probable jet direction, and there
is a circular region of UV and [\ion{O}{2}] emission to the east of the galaxy 
that 
may fall at the boundaries of the eastern lobe (seen in projection on the sky). 
The radio lobe fills the region between the galaxy and the eastern hotspot
but much of the extended emission is resolved out in Fig.~\ref{merlincont}
due to the high MERLIN resolution. 
We will be discussing details of
these central components in the rest of this paper. Also interesting and
indicative of some kind of interaction of the lobe with the ambient
medium is the optical emission directly to the west of the western radio lobe:
this lobe is flattened parallel to this linear structure.

\subsection{Imaging}\label{imageanalysis}
The WFPC2 F622W image and the CFHT [\ion{O}{2}] image have been previously
presented and discussed in RS97,
along with a $K'$-band image obtained with NIRC on Keck I.  
By adding the high resolution NICMOS F160W image
and the F814W image obtained from
the {\em HST} archive, we can determine the morphology of the galaxy and
colors of some components over a large range in wavelength. 
WFPC2 images of 3C280 are also shown and discussed in \citet{bes96,bes98}
and the NICMOS image is shown and discussed in \citet{zir03}.


\begin{figure}[!tb]
\epsscale{0.9}
\plotone{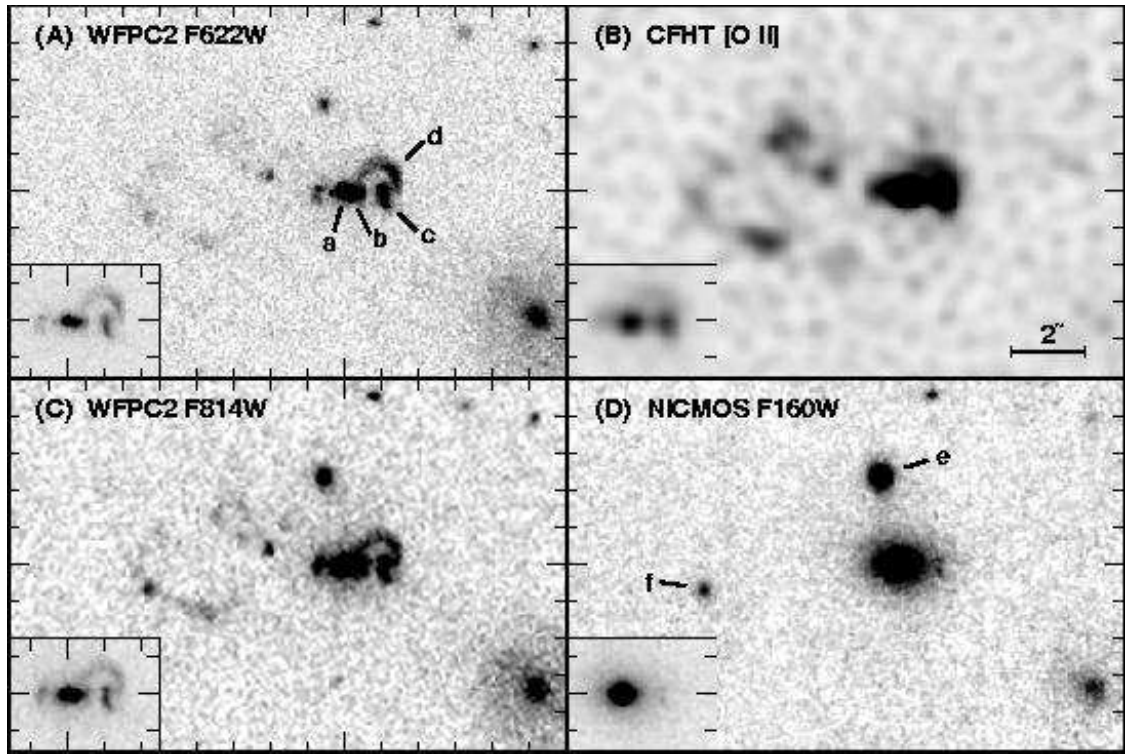}
\caption {High-resolution images of 3C\,280 in different bandpasses.
  The insets show lower-contrast images of the central region.
  {\it (A)}  {\it HST} WFC F622W image (from RS97).
Core components $a$ and $b$ can be seen more clearly in the inset to
Fig. \ref{xbandmap}.
  {\it (B)}  Image obtained with the CFHT SIS fast guider through a
  $\sim30$ \AA\ interference filter centered on the [\ion{O}{2}] $\lambda$3727
  line (from RS97).  The image has been continuum subtracted
  and slightly deconvolved.
  {\it (C)}  {\it HST} WFC F814W image.
  {\it (D)}  {\it HST} NICMOS F160W image.  Note that the two components of
  object c are still visible.}
  \label{images}
\end{figure}


As shown in Fig.~\ref{images}$A$, $C$, and $D$,
an unresolved ``nuclear'' optical component $a$ coincides with
the near-infrared elliptical center; discrete components $b$ and
$c$ are aligned with $a$, on a line very close to the radio jet axis.
In fact, the location of component
$b$ corresponds so closely to the morphology of core jet extension in the X
band map, that it is possible that we are seeing optical
synchrotron in $b$. 
In RS97, three of the 5 $z \sim 1$ quasars we studied with high resolution
HST images had clear evidence for optical synchrotron from jet knots
or lobes. 
For 3C280, we estimate the spectral indices from 8.4 GHz to 6600 \AA\ for the two 
components $a$ and $b$
by fitting overlapping profiles to obtain fluxes in the X band map and
in the F622W optical image (with the underlying emission from
the host galaxy subtracted). The radio-to-optical spectral indices of $a$ and
$b$ are 0.76 and 0.79 respectively, with errors of $\sim$0.2 for $b$, the 
less well-determined component. (Here $F_\nu \propto \nu^{-\alpha}$.)
These spectral index values are reasonable for optical synchrotron emission and
certainly within the wide range possible. 
The radio-to-optical spectral indices for components of 
optical synchrotron jets observed with HST as compiled by \citet{cra93}
range from 0.58 to 0.90, so $b$ would have a typical spectral index were
it due to optical synchrotron emission. 
In addition, the spectral index of component $b$ is not flatter than that of
the core $a$; if it were, it would be difficult to explain it as a component of
the jet. 

Component $c$ is elongated nearly perpendicularly to the radio axis,
and it is apparently connected by a semi-circular arc $d$ to $b$.
The component $c$ is double at longer wavelengths:
in Fig.~\ref{images}$D$ it can be seen that there is a gap
in $c$ that corresponds closely to the jet axis (as determined
from the X band map). 
The WFPC2 F622W filter excludes all
major emission lines at the redshift of 3C\,280, and
these structures are therefore dominated by continuum emission.
(We have estimated the contamination 
from weak emission lines from the spectroscopy presented in \S3.3.2. In 
all cases, this contamination is $\le14$\%.)
The arc has strong [\ion{O}{2}] emission as well, however,
as can be seen in our continuum-subtracted
CFHT [\ion{O}{2}] image, shown in Fig.~\ref{images}$B$. This image had an
initial resolution of 0\farcs7, and its resolution was improved
by deconvolution with the Lucy restoration algorithm.
The background artifacts induced by the deconvolution
process makes it difficult to use the
image for quantitative measurements at the higher resolution,
but the image is reliable enough to
indicate that $c$ and $d$ both have fairly strong
line emission.  They also appear to have roughly similar line-to-continuum
ratios.

The arc $d$
appears to be edge-brightened within an angle of $\sim30\arcdeg$
from the radio axis; this is within the probable quasar opening angle
\citep[\eg][]{sai94,wil94}. It seems likely that this edge brightening
is related in some way to the radio source and could be the
result of illumination by the active nucleus.
One subtle morphological clue is probably best seen in
Fig.~\ref{merlincont}, marked with an arrow: 
there appears to be a faint echo of the arc structure
parallel to the main arc, at a radial distance of about 0\farcs4.
Though very narrow, it seems to be fairly well defined and may well be real,
even though it is not evident in the (shorter exposure) F814W image.

The NICMOS F160W image provides IR imaging at a resolution comparable
to that of the WFPC2 imaging in the optical. 
As was already indicated by the
NIRC $K'$ image, much of the spectacular structure seen 
in the optical images is no longer evident at rest-frame 8000 \AA.
At this wavelength, 3C\,280 is dominated by a quite regular elliptical profile,
though the object $c$ is still visible, along with the probable companion 
galaxy $e$ and continuum object $f$.

Detailed exploration of the radial surface-brightness profile confirms
that the light distribution follows quite closely a de Vaucouleurs
$r^{1/4}$ law and is not well fit by an exponential disk model. We
have taken two different approaches to this fitting.  In both cases,
we use a NICMOS F160W PSF from \citet{rid01} both to subtract off a
point-source component and to convolve with the calculated galaxy models.

The first approach used the fitting procedure described in \citet{lac02},
which iteratively fits a two-dimensional (2D) model to the galaxy, consisting of
a point-source component and an underlying elliptical or spiral galaxy
model. A $\chi^2$ statistic is generated to assess the
goodness of the fit of the model to the data, allowing some estimation of 
the uniqueness of the model fits. 
(Galaxy $e$ and component $c$ were removed from the image before
fitting.) The best-fit model included an unresolved nucleus contributing about
10\% of the total $H$-band flux.  While a model with no point-source
contribution has a $\chi^2$ that is worse by only 1$\sigma$, the 
2D residuals
in the central region are clearly poorer than for the model with a
point-source component.  The best-fit model has a de Vaucouleurs
profile with an effective radius
$r_e=0\farcs59$, an eccentricity of 0.29, and a position angle (PA) of 86\arcdeg

Our second approach was to make a similar two-component,
2D fit by inspecting visually the residuals after subtraction of the model
from the data. 
After an estimated point-source component was
subtracted, we fit the residual elliptical using 
the STSDAS task {\it ellipse} to measure the
surface-brightness profile, varying the eccentricity
and the PA separately to determine a consistent set of parameters.
Using the best average values of the eccentricity and PA
from these preliminary fits,
we then constructed a range of two-dimensional de Vaucouleurs models, varying 
the $r_e$ and the point source contribution, until a best fit to the 
image profile is found.  This
approach gave $r_e=0\farcs76$, an eccentricity of 0.28, and a PA of 85\arcdeg.
The point-source contribution was 9\% of the total flux, corresponding to
an $H$ magnitude of $\sim$21.
The effective radius is similar to that found in
the first method, and the variation in the result gives some
idea of the error in our measurement.

This nuclear
$H$ magnitude corresponds to an absolute V magnitude of $\sim$ $-$21,
which would be a very faint quasar luminosity were the nucleus 
unobscured. However, \citet{don03} find from {\it Chandra} ACIS-S 
that the nucleus of 3C\,280 is a hard source, with an unusually flat X-ray 
spectral index, indicating absorption in the soft X-ray band.
This finding is consistent with the central source of 3C\,280 being an 
obscured, luminous quasar rather than an unusually faint one (compare 
the result for Cygnus A, \citealt{you02}). 

\begin{figure}[!bt]
\epsscale{0.5}
\plotone{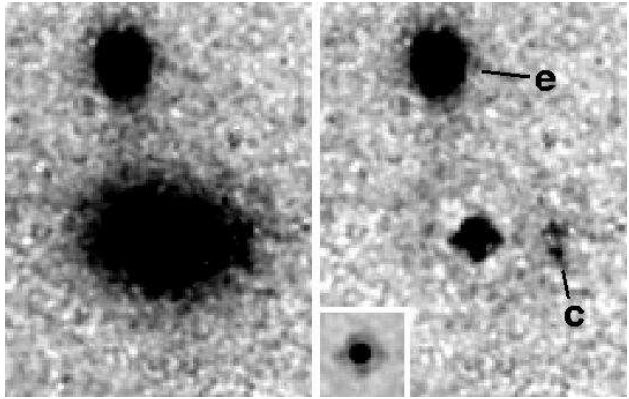}
\figcaption{3C\,280 NICMOS F160W image on the left; on the right, the same
with a best-fit elliptical profile subtracted. The images include the
apparent companion galaxy labelled {\it e}.
An inset shows the residual nuclear contribution at a different contrast. 
For the elliptical
model, the center, ellipticity, and position angle of the major axis
were all held constant. The images are 4\farcs1 $\times $ 5\farcs3;
north is up, east is left. 
\label{nicsub}}
\end{figure}

Figure~\ref{nicsub} shows the
residual after subtraction of the best-fitting elliptical model.
The residual nuclear point source has a FWHM that is roughly consistent
with the one bright star in the field, and it has a clearly visible diffraction ring. 
The rest of the residual map is quite smooth: 
good evidence that some kind of central compact component
is necessary to provide a good fit to the data. 

\begin{figure}[!tb]
\epsscale{0.6}
\plotone{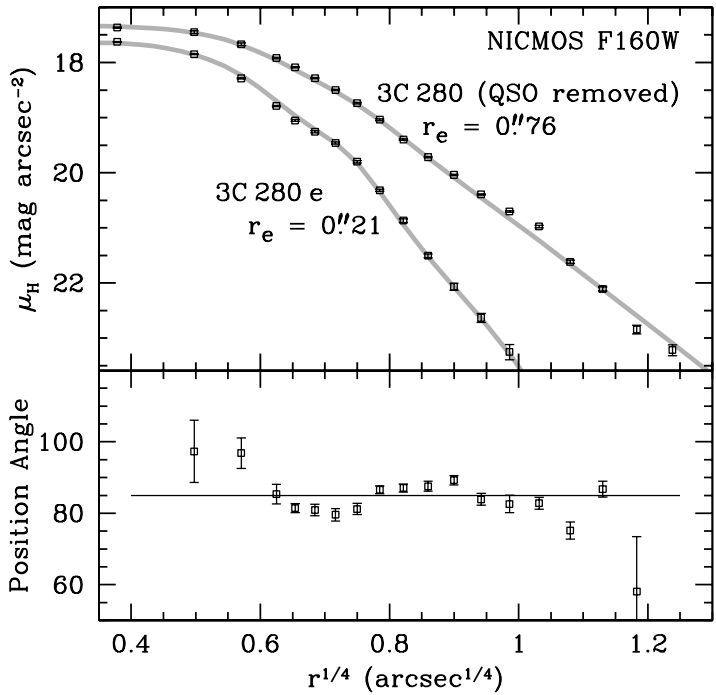}
\figcaption{Upper panel:
Radial-surface-brightness profile for the spheroidal components of
3C\,280 and of the apparent companion $e$, as measured from the NICMOS F160W
image.  The elliptical fit was made assuming a fixed center, eccentricity,
and position angle.
The effective radii are $r_e = 0\farcs76$ and $r_e = 0\farcs21$,
respectively, corresponding to 6.1 kpc and 1.7 kpc.
The gray lines show PSF-convolved $r^{1/4}$-law profiles with these values
for $r_e$.
Lower panel: The position angle variation over the same radial region
for an elliptical fit made to the spheroidal component of 3C\,280
assuming a fixed center and fixed eccentricity.
The position angle of the radio axis (PA = 85\arcdeg) is shown with a 
solid line. 
\label{ellprof}}
\end{figure}

The radial surface-brightness profiles for both 3C\,280 and the apparent
elliptical companion $e$, along with the best-fit models, are shown in
Fig.~\ref{ellprof}.  The bottom panel in Fig.~\ref{ellprof} shows the
fit to the observed profile with the center and eccentricity held fixed,
but with the PA allowed to vary.  The errors become large at very small
and very large radii, but the PA is remarkably constant over most of the
range. This consistency demonstrates clearly that the close alignment of
the elliptical PA with the radio axis is found at all radii.

At $K$, with our ground-based $\sim$0\farcs7 seeing, we also found 
the galaxy profile to be well fit by an $r^{1/4}$ profile with 
an $r_e$ of $\sim$0\farcs85 and fixed PA = 85\arcdeg, with an estimated
nuclear contribution to the elliptical of $<$10\% (RS 97).  
The higher resolution data has confirmed what could not be shown 
with ground-based seeing: it is in fact the major axis
of the regular elliptical galaxy that appears to be aligned with the radio jet,
and the result is not due to a discrete aligned contribution that was 
unresolved at ground-based resolution. 
This is further borne out by inspection of the residual image for our
elliptical fit. The double components
of object c were clearly ignored by the fitting procedure,
which takes the median flux along the elliptical isophote and is therefore
insensitive to small, discrete features.
The smooth residual (except for the nucleus and $c$) shows that there
are no extra discrete components that could have twisted the
elliptical isophotes into alignment with the radio axis; such features
would show up as structure in the residual map. 
Any of the aligned UV components which dominate in the WFPC2 image 
would have been resolved in the NICMOS image and therefore would 
have affected the structure of the residual map, and 
thus they cannot account for the alignment of the elliptical fit. 
We cannot distinguish in the image between stellar and non-stellar
light, and therefore we cannot rule out the existence of a broader,
red aligned component with a profile similar enough to the stellar
profile to give the sum of the components a good de Vaucouleurs fit.
The spectrocopy, however, indicates that the stellar
population should be fairly dominant at rest-frame $\lambda$ $>$ 4000\AA,
and we see no evidence for a non-stellar red component. 


.


\subsection{Spectroscopy}\label{specanalysis}

\subsubsection{Analysis of the spectra}

\begin{figure}[!tb]
\epsscale{0.9}
\plotone{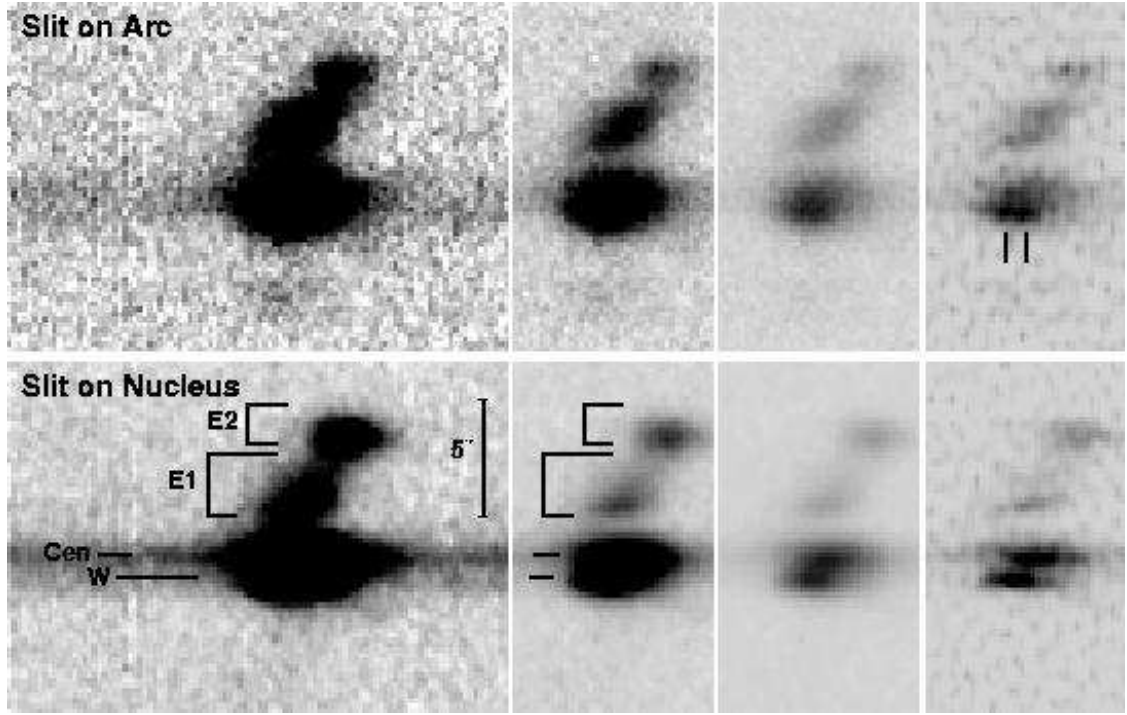}
\figcaption{Two-dimensional spectral images of the [\ion{O}{2}] $\lambda3727$
doublet for the two slit positions shown in Fig.~\ref{slitfig}.  The first
three panels (left to right) for each slit position show successively
lower-contrast versions of the line profile; the fourth panel shows a
deconvolution of the profile.  The locations of regions for which we have
extracted 1-d spectra are shown.  The two short vertical lines in the
upper-right panel indicate the separation of the two components of the
[\ion{O}{2}] doublet.
\label{oii2dfig}}
\end{figure}

In Fig.~\ref{oii2dfig} we show extracts of the two-dimensional [\ion{O}{2}]
line emission from both of our slit positions. 
Continuum is obviously visible from the nuclear region and
from a distinct component to the west of the nucleus, 
and there are several distinct components to the east with
varying velocities.
We have named these primary components according to
their relative position with respect to the nucleus: $W$, $Cen$, $E1$, and $E2$,
and we show these labels next to the [\ion{O}{2}] profile for the nuclear slit 
as well as on the slit map (Fig.~\ref{slitfig}).

In order to measure fluxes for the different physical
components, we have extracted one dimensional spectra parallel to the
central continuum with aperture widths of 4 pixels,
corresponding to about one spatial FWHM (0\farcs86). 
The components $E1$ and $E2$ extend over several 
FWHM, while $W$ is confined to one FWHM. 
We refer to the subdivisions of $E1$ and $E2$ as $E1a$, $E1b$,
$E1c$, $E2a$, and $E2c$ when we wish to preserve our
best spatial resolution, but we  
sum these individual spectra when discussing the overall physical properties
of components $E1$ and $E2$.
The central spectrum has been extracted with a slightly
wider aperture of 5 pixels (1\farcs08) around the central
continuum to minimize other
contributions while maximizing signal-to-noise.
It can be seen that the spectral extraction
$W$ is dominated by the imaging component $c$, while 
the spectral extraction {\it Cen} corresponds to the components
$a$ + $b$. The spectral extractions $E1$ and $E2$ include
portions of the eastern emission at differing distances from the nucleus.
In Fig.~\ref{cenwestsp} we show line identifications for all lines
seen in the central spectrum and in spectral component W.


We have measured fluxes in the various 1-D spectra
with Gaussian fits (multicomponent, where necessary due to line overlap) and
give the results for the central components ({\it Cen} and $W$)
in Table \ref{tabcen} and for the (sub-divided) 
eastern components ($E1$ and $E2$) in
Table \ref{tabeast}.
Wavelength errors are $\sim$ 1 \AA, and statistical errors in the fluxes
are $\sim$ 10\%, except where noted in the table. In some cases of
very faint lines, fluxes are
measured after smoothing the spectrum with a 5 pixel ($=6.3$ \AA) kernel.

\begin{figure}[ht]

\plotone{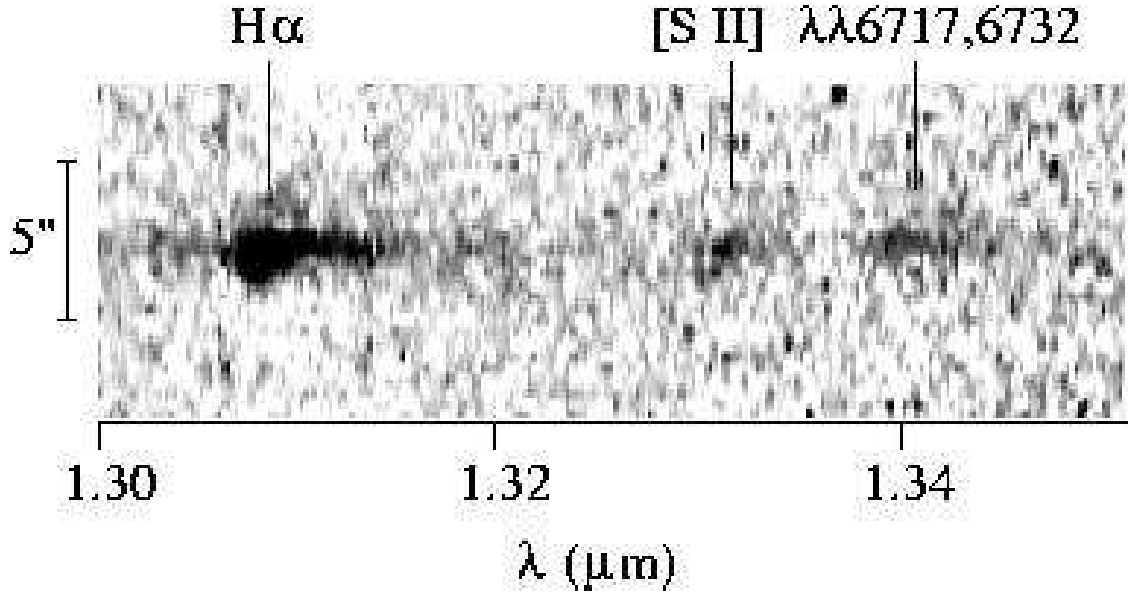}
\figcaption{The long wavelength $J$ band CGS4 spectrum with the primary emission
lines marked. \label{jbandspec}}
\end{figure}

We show the long wavelength 
$J$ band near-infrared spectrum in Fig. \ref{jbandspec}.
To allow comparison of the lines detected in the near-infrared with
those in the optical spectrum,
we have transformed the near-infrared spectra
centered on the 3C\,280 nucleus to match the optical spectra in
spatial scale.
We have extracted 1-D spectra at the positions expected for the
primary extended line components seen in the optical, with the larger
apertures necessary due to the poorer spatial resolution of the near-IR
spectra. We can compare these to matching extractions made from the central
optical spectrum. Seeing conditions in both the optical
and near-IR observing runs were fairly well matched to the slit size, and
it can be seen from the [\ion{O}{2}] image that the central line components 
are fairly compact. However, without any direct spectral overlap it is difficult
to account for relative flux calibration, and we might expect errors
as much as 30--50\% between the different spectra. 
We have measured the near-IR line fluxes with one component Gaussian fits,
fitting the continuum directly in the region of the line, and give these in 
Table \ref{irlinetab}. Fluxes were measured after smoothing with a
5 pixel Gaussian (except for the $H\alpha$ flux).


\subsubsection{Observed Emission Line Spectra}

\begin{figure}[!tb]
\epsscale{0.9}
\plotone{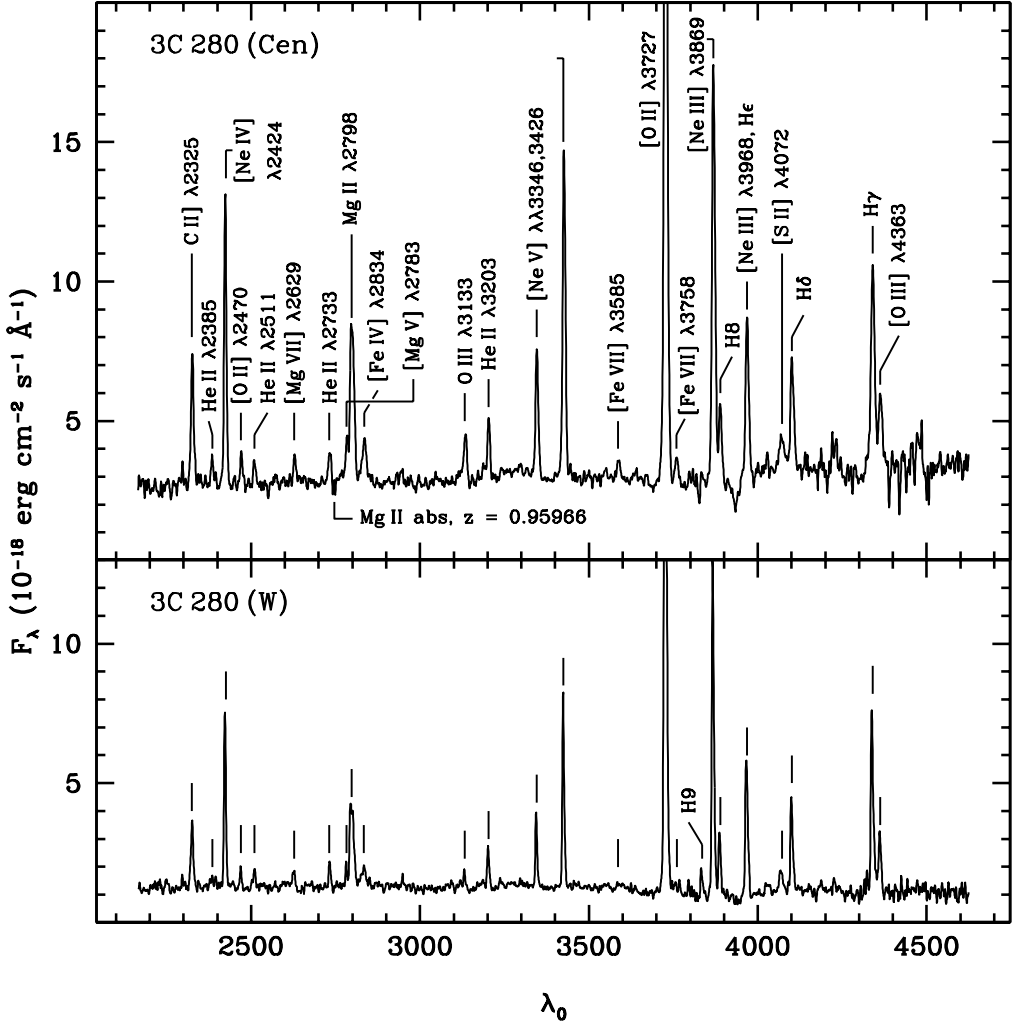}
\figcaption[cenwestsp.eps]{Spectra of the central ($a+b$) and western ($c$)
components.  Emission lines are identified in the upper panel, and the
same tick marks are reproduced in the lower panel
(even if no line is present).  Note the presence of \ion{Ca}{2} absorption
in the central component between H8 and [\ion{Ne}{3}] $\lambda3968$ +
H$\epsilon$.\label{cenwestsp}}
\end{figure}

\begin{figure}[!tb]
\epsscale{0.9}
\plotone{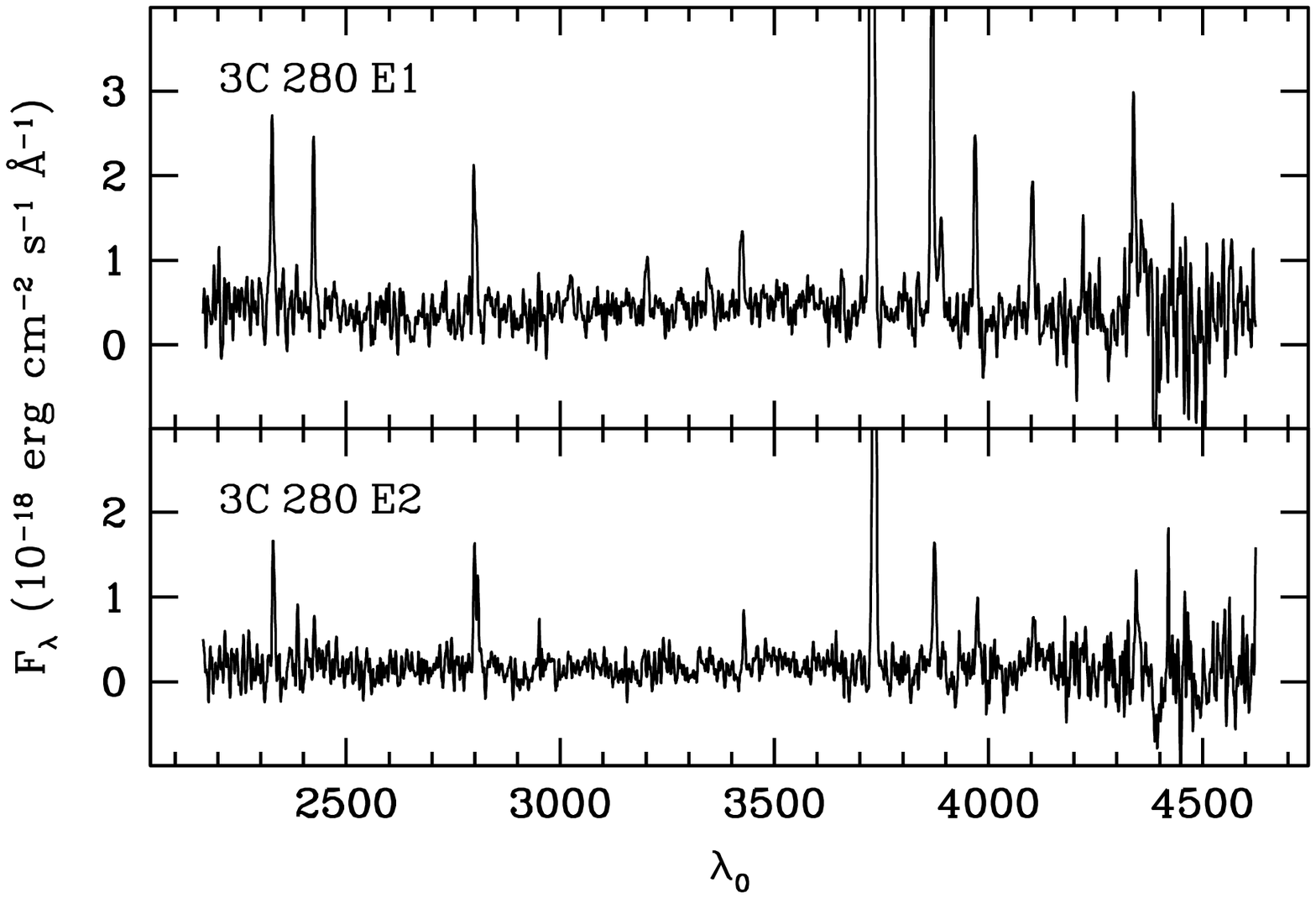}
\figcaption[eastsp.eps]{Spectra of 2 regions in the vicinity of the eastern
radio lobe; the positions of these regions on the spectra
are shown in Fig.\ref{slitfig}.\label{eastsp}}
\end{figure}

The spectra of the central and western components are shown in 
Fig.~\ref{cenwestsp} and of the eastern components in Fig.~\ref{eastsp}.  
The central and western optical spectra have good S/N for a wide range of
emission lines.
There is also a strong stellar component, as 
the stellar absorption feature \ion{Ca}{2} K and a clear 4000 \AA\
break are visible directly.  
In addition all regions show continuum emission, although
little from the eastern components E1 and E2.  
The emission lines ratios vary from region to region.


First, we check from the Balmer line ratios in the optical spectrum
the reddening to the emission line gas:
from $H8$ to $H\gamma$ in the central spectrum and from $H10$ to $H\gamma$
in component W we find that the reddening is probably reasonably low,
as the ratios are basically consistent with Case B recombination
(Osterbrock 1989). 
We therefore assume reddening to be zero throughout the rest of the discussion.

One straightforward measure of the ionization properties in
gas is the [\ion{Ne}{3}]$\lambda$3869/[\ion{Ne}{5}]$\lambda$3426 ratio 
\citep{ost89}. 
The relative strengths of these lines are determined by the 
ionization parameter, assuming temperature and electron density to be 
similar. 
We use this line ratio instead of the [\ion{O}{2}]/[\ion{O}{3}] line ratio
commonly used in studies of low-$z$ objects, since
the [\ion{Ne}{3}] to [\ion{Ne}{5}] line ratio will be little
affected by reddening, and it doesn't require us to use the infrared spectrum
(for which the calibration relative to the optical has uncertainties). 
We find that the line ratio is lowest in the central component (ratio=1.5) 
and generally increases with distance from the nucleus where 
well-measured. 
The line ratio in $W$ is 2.1, while in the brightest 
sub-components of $E1$ and $E2$ ($E1a$ and $E2a$) the ratios are 3.5 and 5.3,
respectively. The line ratio therefore increases
with distance in these eastern components. 
This decrease in the ionization state is consistent
with photoionization from the central source being the primary
mechanism heating the emission line gas, although 
density effects may also play
a role in the differences among these spectra.

\citet{bes00b} used the intensity ratios of \ion{C}{3}] 
$\lambda1909$/\ion{C}{2}] $\lambda2326$ and [\ion{Ne}{3}] 
$\lambda3869$/[\ion{Ne}{5}] $\lambda3426$ to infer whether the 
ionization of the gas was primarily due to shocks or photoionization in 
9 $z\sim1$ radio galaxies, including 3C\,280, which they conclude is 
within the photoionized domain. Our spectra do not include \ion{C}{3}] 
$\lambda1909$, so we cannot construct similar diagrams for the various 
individual components of 3C\,280. While the \ion{C}{3}] 
$\lambda1909$/\ion{C}{2}] $\lambda2326$ ratio used by \citet{bes00b} 
was from their own recent spectroscopy, the
[\ion{Ne}{3}] $\lambda3869$/[\ion{Ne}{5}] $\lambda3426$ ratio of 0.7 
that they used was from early spectroscopy by \citet{spi82}. Our larger 
values for this ratio in 3C\,280 actually give even better agreement 
with the photoionization models, provided that the
\ion{C}{3}] $\lambda1909$/\ion{C}{2}] $\lambda2326$ ratio stays close 
to that found by \citet{bes00b}. This is almost certainly the case for 
the central region, which dominates the
emission-line spectrum, so we can fairly definitely say that the gas 
there is photoionized. The upper limit of $\sim0.1$ to the \ion{C}{1}] 
$\lambda2966$/\ion{C}{2}] $\lambda2326$ ratio reinforces this 
conclusion: for the shock models of \citet{dop96}, this ratio is 
greater than this in almost all cases except for very high magnetic 
parameters.

Another emission line diagnostic is the relative strength of the 
[\ion{O}{3}] $\lambda\lambda4959$,5007 and [\ion{O}{3}] $\lambda4363$ lines. This 
is a temperature diagnostic, and can be used as an indicator
of whether emission line gas properties are more consistent with shock
heating or photoionization as the primary excitation mechanism.
In our near-infrared spectra we were able to detect $\lambda$5007 
in the central region, $W$, $E1$ and $E2$. (In the central region, we
detected also $\lambda$4959 with the expected flux of about one 
third $\lambda$5007).
In the optical spectra, [\ion{O}{3}]~$\lambda$4363 was detectable only in the
central component and $W$, not in the extended emission line gas to the east,
$E1$ and $E2$. The ratio of [\ion{O}{3}]~$\lambda\lambda4959$,5007/[\ion{O}{3}]~$\lambda4363$
is 110 in the nucleus and about 95 in the spectral component $W$ (which includes
primarily emission from the component $c$ seen in Fig. \ref{images}).
The relative calibration between the near-IR and optical
spectrum could be incorrect by some 30\%.
In any  case, the line ratios are clearly quite high, and unlikely to be 
consistent with the values of 10--20 predicted for pure shock ionization
models \citep{sol01,dop96}. 
The ratios we find are comparable to or less than 
those predicted by pure photoionization models or those 
found in low-$z$ radio galaxies considered to be primarily photoionized 
\citep{sol01}, while somewhat higher than that found for 
4 galaxies (ratios $\sim$50) 
that had significant kinematic signatures of shocks.
The ratios we find are therefore basically consistent with photoionization
being the dominant excitation mechanism in the gas, but not inconsistent with
mixed models that include precursor ionization contributions from shocks.

In each individual region, the gas seems fairly quiescent, with no particularly
large linewidths. In the arc spectrum, the [\ion{O}{2}]~$\lambda3727$
doublet is resolvable throughout the extended emission line region.
As can be seen from Fig. \ref{oii2dfig}, the line centers in the
eastern components are redshifted, with the farthest component $E2$ 
having a larger relative velocity. Internal to component $E1$, some kinematic
signature of slightly increasing velocity with distance is also
visible (particularly in the upper spectrum showing the slit placed on the 
arc position). 
These eastern emission line regions surround part of the eastern radio
lobe, and the redshifted velocity gradient is consistent (for the 
orientation of the object) with this
gas being pushed out by the radio lobe. It is surprising
that the velocities are not higher, however, if the material 
has been directly entrained in the lobe. 
The emission line gas in component $W$ (mostly image component $c$)
also is consistent with the probable orientation of the
object, in that it exhibits a slight blueshift.
Both emission line region velocities are consistent 
with outflow along the jet direction. 
This is also a kinematic signature of rotation, although the rotation
axis would have to be anti-correlated with the angular momentum 
axis of the black hole producing the jet.
\citet{bes00a}  suggest that
a rotation of the halo and the host galaxy
could explain some of the kinematic signatures
in the emission line gas they studied around $z \sim 1 $-- 2 3C radio
galaxies.  
In general, 
the total velocity shifts and widths
are moderate, at most a few hundred km s$^{-1}$;
there is nothing similar to the thousands of km s$^{-1}$ shifts
seen in radio galaxies like 3C 368, where radio-jet interactions causing 
shocks must be important.

\subsubsection{Decomposition of the spectra }

We have so far discussed only the emission line properties of the gas,
but we have clearly detected significant continuum in the central
components, as well as stellar absorption lines. We therefore wish
to separate, if possible, the different spectral components which
contribute to our observed spectra. We will discuss each
spectrum in turn. 

\begin{figure}[p]
\epsscale{0.75}
\plotone{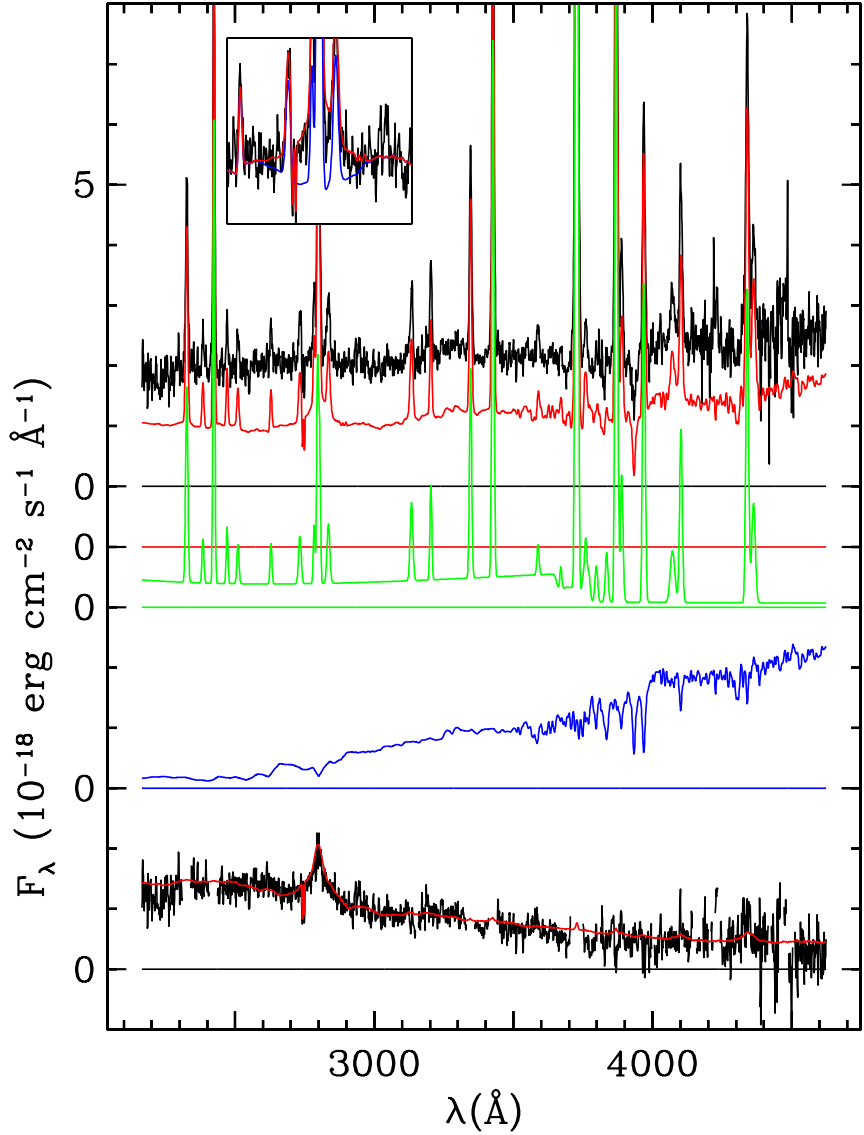}
\figcaption[cencomponents.eps]{Modeling the spectral components contributing
to the continuum of the central ($a+b$) object.  The black trace at the top
reproduces the spectrum of the top panel of Fig.~\ref{cenwestsp} at an
enlarged vertical scale.  The red trace shows the model spectrum obtained
by summing the three components given below:  emission lines and nebular
thermal emission (green), a \cite{bru96} 3-Gyr-old stellar population
model (blue), and a QSO spectral model (red; superposed on the residual of
subtracting the emission-line and stellar models from the observed spectrum).
Baselines are offset as indicated for clarity.  The inset in the upper left part
of the plot shows the \ion{Mg}{2} region in more detail: the black trace
is the observed spectrum, the red trace is the model sum, and the blue
trace is the model sum with the broad \ion{Mg}{2} profile removed from the
quasar model. 
\label{cencomponents}}
\end{figure}

\subsubsubsection{The Central Component}
Figure~\ref{cencomponents} shows an attempt to isolate the various contributions
to the observed continuum for the central component.  We start by subtracting
the emission lines and the nebular thermal continuum (\ie\ free-bound,
free-free, and two-photon).  For the latter, we have assumed Case B for the
hydrogen emission, a temperature of 15000 K (except as noted), an average
ionization fraction of 0.1 for helium, and the low-density
limit for two-photon emission.  
For most of the lines, we simply have fitted
a Gaussian profile assuming an interpolated continuum for a baseline.
However, some lines are affected by
features in other components; for these, an iterative approach was required
to arrive at a self-consistent solution.  The lines affected include the
Balmer emission lines, the [\ion{Ne}{3}] $\lambda\lambda3869$,3968 lines, and
lines in the vicinity of the \ion{Mg}{2} $\lambda\lambda2796$,2803 doublet.

Once we have fixed the fluxes for the H$\gamma$ and H$\delta$ lines, the
scaling of the thermal emission is determined.  The [\ion{Ne}{3}] $\lambda3968$
emission is confused by H$\epsilon$ emission and absorption and by
\ion{Ca}{2} H absorption, but its ratio of 1/3 (in photon units)
to the stronger [\ion{Ne}{3}] $\lambda3869$ line allows it to be subtracted
quite accurately.

The observed spectrum clearly shows the \ion{Ca}{2} K absorption line
and evidence for a 4000 \AA\ break, the latter, at least, indicating the
presence of a relatively old stellar population. The presence of old stars
can also be inferred from the closely $r^{1/4}$-law radial-surface-brightness
profile at rest-frame wavelengths of 1 \micron\ (RS97) and 0.8 \micron\
(\S\ref{imageanalysis}).
However, no single or composite stellar population could be found that
would adequately reproduce the entire residual continuum, and we
also find a broad component to the \ion{Mg}{2}
$\lambda2798$ emission, indicating direct or scattered emission from a
quasar nucleus.

To isolate the contribution from the stellar component, 
we have tried to find one that would leave as smooth a residual as possible.
We have experimented with stellar models
comprising various combinations of Bruzual \& Charlot (1996) isochrone synthesis
stellar populations (restricted to the higher-resolution, solar-metallicity
ones that roughly match the resolution of our data).  We rapidly converged
towards a dominant contribution from a population older than about 1.5 Gyr,
necessary to provide the observed curvature in the spectrum between about
2800 and 3600 \AA, while also fitting the 4000 \AA\ break and absorption
features at longer wavelengths.  However, populations older than about
3 Gyr, which otherwise fit well, 
oversubtract the flux at the long-wavelength end. 
With the S/N of this composite spectrum, 
we cannot rule out small admixtures of a younger population
but no
significant portion of the light can come from populations with ages of
50 Myr or younger because of constraints on the residual flux at short
wavelengths.  

In the following subsection, we attempt to isolate the elliptical component,
and there derive a stellar population of 2--4 Gyr.
Therefore, for simplicity, 
in Fig.\ \ref{cencomponents}, we have used a single,
instantaneous burst population with an age of 3 Gyr, which fits as well
as any of the composite populations we have tried.  This model should be
regarded as a proxy for any more complex star-formation histories that could
give similar SEDs.  

The inset in Fig.~\ref{cencomponents} shows the region around the
\ion{Mg}{2} line in more detail, including a blue trace showing the
consequence of omitting the broad \ion{Mg}{2} profile from the QSO
model spectrum. The appearance of ``oversubtraction'' around the
\ion{Mg}{2} line is wholly due to the dip in the stellar population
model at that wavelength. The broad \ion{Mg}{2} profile itself is well
fit by a Lorentz profile with a rest-frame FWHM of 56 \AA. It is not
clear whether we are seeing the broad-line region directly or via 
scattering by dust or electrons. If we are seeing it directly, and the
broad-line region has an extinction similar to that inferred for the
QSO continuum from the hard x-ray spectrum \citep{don03}, only a very
small percentage of the lines of sight can be unobscured. If the scattering
is by dust, particularly (as would seem likely) optically thick dust, there
are a number of combinations of scattering parameters that would
preserve the approximately gray scattering that we seem to see over
our restricted wavelength range (see, \eg\ \citealt{zub00}). If the 
scattering is by electrons,
the lack of substantial broadening beyond the typical QSO
profile is inconsistent with the scattering electrons having a
temperature $> 10^5$ K (see, \eg\ \citealt{cim96}).
 
There are several cases among $z\sim1$ radio galaxies for which
spectropolarimetric observations have detected broad \ion{Mg}{2} lines
\citep{dey96,cim96,cim97}, and sometimes this broad emission is
discernible in the total flux as well. The example most similar to our
observations of 3C\,280 is that of 3C\,356 \citep{cim97}, for which the
broad \ion{Mg}{2} is detected in the apparent nuclear component.
Whereas we obtain a total flux of $7.5\times10^{-17}$ erg cm$^{-2}$
s$^{-1}$ for the broad component, corresponding to a luminosity of
$3.9\times10^{41}$ erg
s$^{-1}$, \citet{cim97} find a flux for 3C\,356 corresponding to a
luminosity of $8.2\times10^{41}$ erg s$^{-1}$ (for our assumed
cosmological parameters). Given that their slit width was twice ours,
these values are identical within the uncertainties.
When we subtract both the emission-line and stellar models from the observed
spectrum, we are left with the black trace at the bottom of
Fig.~\ref{cencomponents}, which shows a continuum that decreases with
wavelength, with a superposed broad \ion{Mg}{2} profile (the latter has
a displaced \ion{Mg}{2} absorption doublet). We have made a composite
QSO SED by smoothly joining, just shortward of the \ion{Mg}{2} profile,
the \citet{zhe97} composite spectrum (which we take
to be more representative of radio-loud quasars, but which does not
extend shortward of 3000 \AA) to that of \citet{fra91}.  We have simply
scaled this spectrum to our residual near 3200 \AA, and aside from adding the
\ion{Mg}{2} absorption profile, we have made no other adjustments.
This composite QSO spectrum is a good fit to our residual
spectrum.  We add this QSO model to our stellar and emission-line models
to obtain the model for our observed spectrum, shown in red just under
the observed data at the top of Fig.~\ref{cencomponents}.  Note that
at about 3000 \AA, the quasar, the stellar component, and the thermal
continuum all make roughly equal contributions to the total continuum
observed in our slit.

Another source of emission is also possible 
in this central component. The close alignment of
the radio map with component $b$ indicates that much of
the F622W flux in $b$ could be optical synchrotron emission. 
If we assume that a synchrotron component with spectral index 1.3 is also
contributing to the central spectrum, we can make a good
model fit by adding a 25\% contribution from 
the synchrotron component and by decreasing the quasar contribution by
$\sim$50\% and the
old stellar population by $\sim$10\% (these percentages
are referenced to the F622W filter). 
This model fit is essentially as good 
as the one without optical synchrotron.

\begin{figure}[p]
\epsscale{0.75}
\plotone{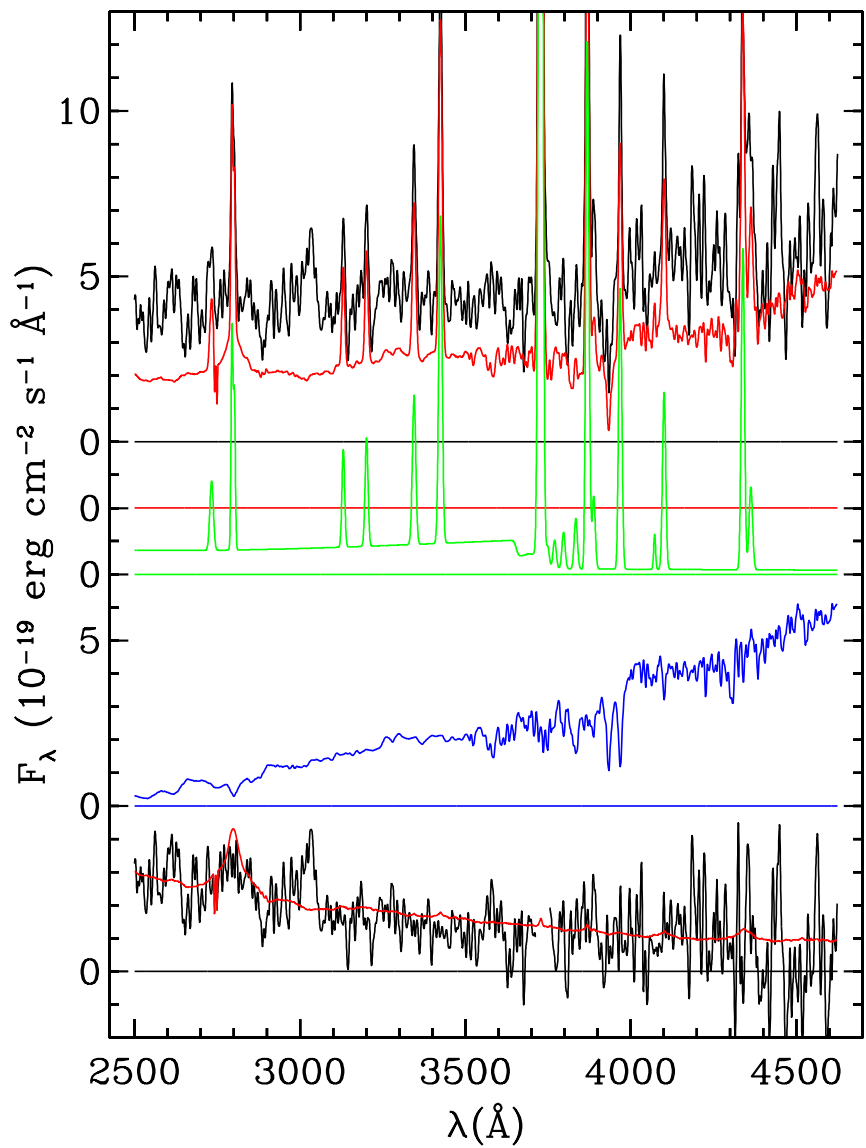}
\figcaption{The observed spectrum of a region $\sim1\farcs3$ east of the
nucleus (a relatively clean observation of the 
underlying elliptical) is shown in black at the top. The red trace immediately below shows
the model spectrum obtained
by summing the three components given below:  emission lines and nebular
thermal emission (green), a \cite{bru96} 3-Gyr-old stellar population
model (blue), and a QSO spectral model (red; superposed on the residual of
subtracting the emission-line and stellar models from the observed spectrum.
Baselines are offset as indicated for clarity.
\label{clrsp}}
\end{figure}

\subsubsubsection{The Elliptical Component}\label{ellipdecom}
We have carried out a similar analysis on the spectrum extracted from
a region $\sim1\farcs3$ east of the
nucleus that is reasonably clear of observed structure in the {\em HST}
WFPC2 images.  Because of its proximity to the nucleus, this region
gives us our best hope of obtaining a relatively uncontaminated spectrum
of the underlying elliptical component dominating the IR images.  The
spectrum obtained for this region is shown in Fig.~\ref{clrsp}.  Although
there is still substantial emission in this region, the stellar component
dominates the continuum longward of about 3700 \AA.
The age of the stellar population is not well determined:  any
solar-metallicity models from $\sim2$ Gyr to $\sim4$ Gyr could be fitted
fairly easily, as could somewhat younger populations with higher metallicities.

\subsubsubsection{The Western Component and the Arc}

We have applied this same spectral decomposition procedure to the
spectra of the western component W ($c$ in Fig.~\ref{images}) 
and of the arc ($d$).  In both
cases, we assume a contaminating stellar population from the
spheroidal component similar to that found above from the ``clear''
region east of the nucleus.

\begin{figure}[p]
\epsscale{0.75}
\plotone{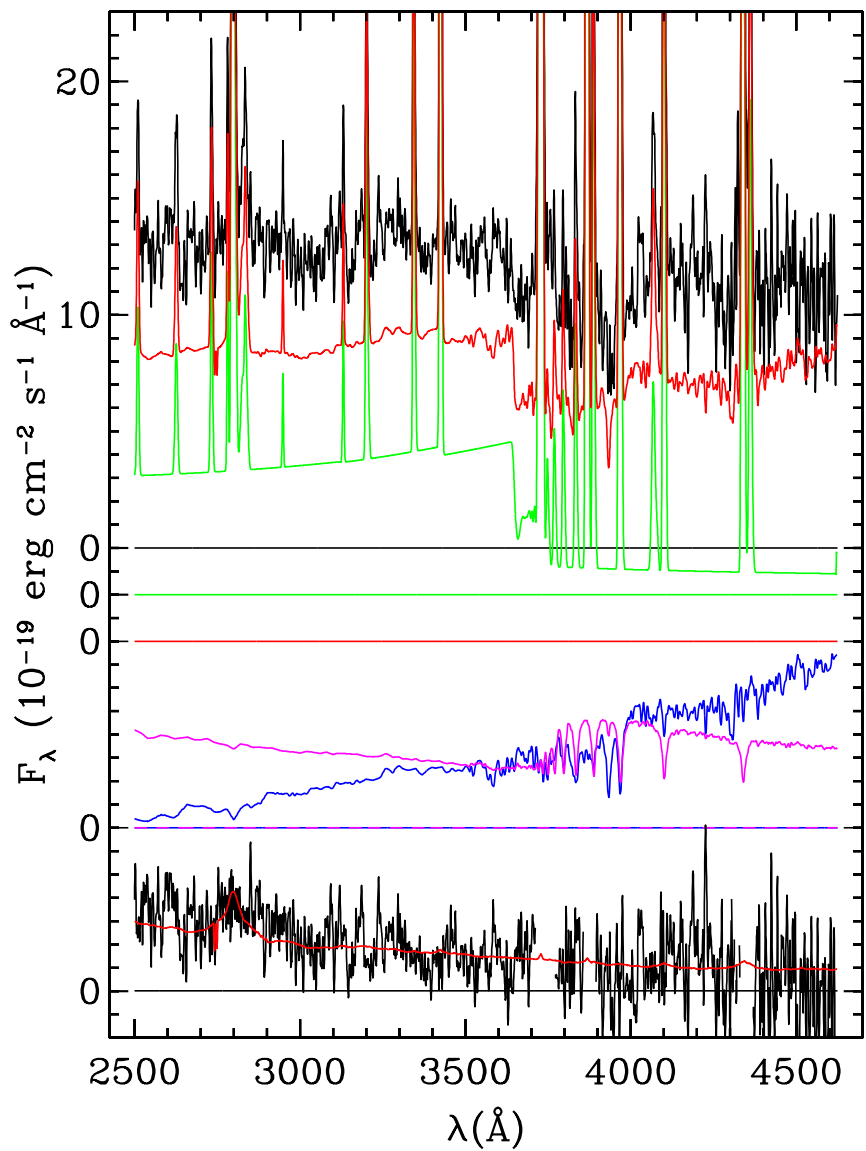}
\figcaption{The observed spectrum of component $c$ is shown as the upper
black trace. The red trace below this shows the model spectrum obtained
by summing the four components given below:  emission lines and nebular
thermal emission (green), a \cite{bru96} 3-Gyr-old stellar population
model (blue), a 0.1-Gyr-old stellar population model (magenta), and a
QSO spectral model (red; superposed on the residual of
subtracting the emission-line and stellar models from the observed spectrum.
Baselines are offset as indicated for clarity.
\label{csp}}
\end{figure}

The spectrum of component $c$ shows relatively strong emission
(Fig.~\ref{csp}). In this
case it was necessary to assume a lower $T_e$ of $10^4$ K for the nebular
thermal continuum in order to achieve a satisfactory fit; this lower
temperature is not unreasonable, given clouds with densities similar to
those in the central component at a greater distance from the ionizing
source.


After we subtract off the emission component, we are constrained by
purely spectroscopic considerations in the amount of 3-Gyr-old stellar
light from the spheroidal component we can subtract.  If we try using
{\it only} this one stellar population, we are left either with an
oversubtraction of the 4000 \AA\ break or with a significant residual
with a fairly flat (in $F_{\lambda}$) SED.  The best solution we have
found incorporates an additional stellar component with an age of
$\sim0.1$ Gyr.  The slightly sloping residual remaining can be
accomodated by our standard QSO spectrum without violating constraints
on broad \ion{Mg}{2} emission.  However, if we were to try to accommodate
enough scattered QSO light to avoid the need for the young stellar population,
we would run into problems with the lack of broad \ion{Mg}{2} emission.
We could wash out this feature with scattering by hot electrons, but
this would require electrons with velocity dispersions $>2100$
km s$^{-1}$. Electrons at this temperature would not be confined to
the relatively small discrete components seen at long wavelengths for
object $c$.

As a check on this model, we can estimate the relative flux expected from
the spheroidal component at the position of our ``clear'' region east of
the nucleus and at the position of component $c$ from the NICMOS F160W image
and compare this ratio to that for our scaling of the 3 Gyr populations in
these two regions.  
Our best estimate gives a
ratio of about 1.4 (for the region centered on $c$ to the ``clear'' region)
from the NICMOS
image, compared with a ratio of scaling factors for the 3 Gyr population in
the spectral decompositions of 1.25.  These are identical within the errors.

Once again, we cannot claim uniqueness for this decomposition.  However,
if the only important contributors to the observed spectrum are nebular
emission, stars, and scattered quasar light, we do have reasonable evidence
for a moderately young stellar population contributing to object $c$,
although well over half of the light at rest-frame 3500 \AA\ (leaving
out the spheroidal component) is due to nebular thermal continuum.
It seems quite likely that the two peaks seen in the longer wavelength
images of $c$ (see Fig.~\ref{images}) are regions dominated by younger
stars (with average ages around 100 Myr),
whereas the more uniform and diffuse appearance of $c$ at shorter wavelengths
shows mostly the distribution of the emission-line gas and scattered
quasar continuum.

\begin{figure}[!tb]
\epsscale{0.75}
\plotone{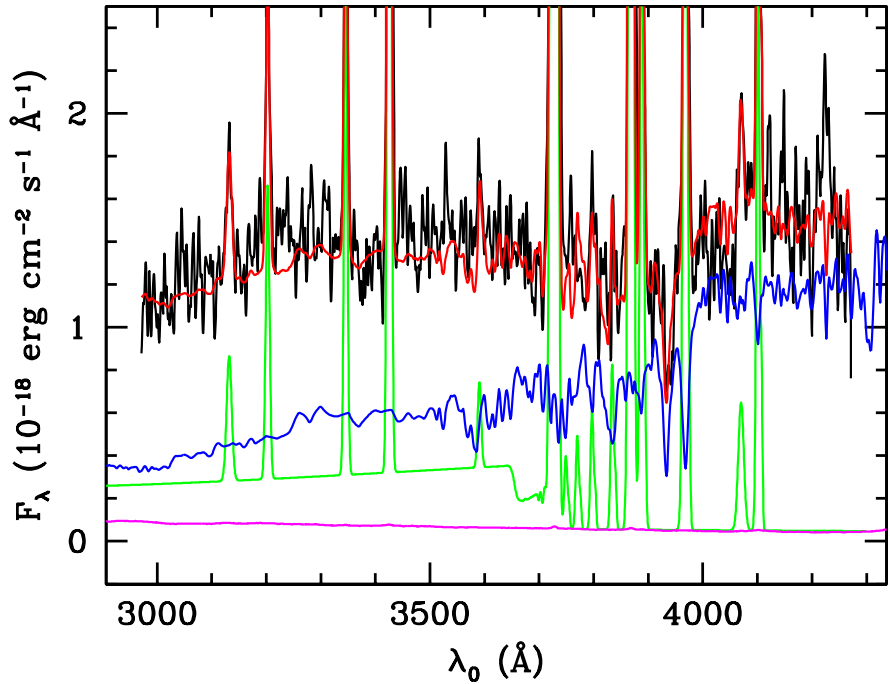}
\figcaption{The observed spectrum of the arc ($d$) is shown as the
black trace. The superposed red trace shows the model spectrum obtained
by summing the three components given below:  emission lines and nebular
thermal emission (green), a \cite{bru96} 3-Gyr-old stellar population
model (blue), and a QSO spectral model (magenta).
\label{arcsp}}
\end{figure}

Our data on the spectrum of the arc $d$ is more restricted in wavelength,
but the decomposition is well constrained.
Careful measurement of relative positions along the slit indicates that we
were successful in obtaining a spectrum of the arc, with little contamination
from other components.  The spectrum is shown in Fig.~\ref{arcsp}.
The nebular thermal continuum is well determined by the strong H$\delta$
line, and the scaling of the spheroidal component is constrained both by
the \ion{Ca}{2} K absorption and the overall shape of the continuum.  We
have experimented with adding in a younger population that might be
intrinsic to the arc, but there is no room for any significant
stellar population younger than $\sim1$ Gyr.  This finding is consistent
with the invisibility of the arc in the NICMOS F160W image.


\subsection{Properties and Analysis of the Stellar Components}

Our spectroscopic analysis has indicated the presence of stars 
due primarily to two components: the underlying elliptical
and an additional younger stellar population in the object $c$.
Using the high-resolution near-infrared imaging, we can check
whether the broad band colors of these components are consistent
with the spectroscopic determination of the ages of the stellar
components. 
We can estimate the elliptical contribution to the F622W band by 
taking the elliptical model from the best-fit to the infrared and
(convolving with the appropriate PSF) estimating the elliptical
contribution to the F622W band. From RS97, we found the optical--IR colors
were consistent with a 2 -- 4 Gyr old stellar population; the 
F160W flux from the elliptical agrees with this earlier result.

This is more difficult for the younger stellar population we infer
to be present in component $c$. 
The two discrete peaks visible in the F160W NICMOS image are most
likely associated with this stellar component, while in the
bluer passbands the contribution from the thermal continuum is dominating.
To estimate the optical--IR colors of this stellar component alone, we
have first measured the total flux in the F622W image that contributed
to spectrum W (in the 1\arcsec\ slit with aperture width of 0\farcs86),
which is dominated by component $c$. From our spectral decomposition
shown in Fig. \ref{csp} we estimate that this young stellar
population should contribute $\sim$25\% of the flux over the rest-frame
passband of the F622W filter. Seeing was good during these observations
and most of the flux seen in the F622W image should have been confined
within the spectral aperture.
We correct the 
F622W flux by this factor and use this value as the blue photometry 
point. The stellar components should be the primary contributors
to the flux in the F160W filter (at rest-frame $\lambda$ 
$\sim0.8 \mu$m). We take the elliptical-subtracted
F160W image (as shown in Fig.~\ref{nicsub}) and measure the flux
in the same aperture (which includes the double peaked
component  $c$ and some background). The ratio of the corrected
F622W flux to the F160W flux for the young stellar component of $c$
is 3.3, but uncertainties in how much flux
of each component would be included in the slit could give ratios
different by a factor of two.
Nonetheless, the flux ratio between these two
filters predicted for a stellar population of 100 Myr is 2.8; 
our broadband photometry is therefore quite consistent with 
the ages of the stellar populations we derived from the spectroscopy. 

In order to assess the mass of the elliptical component in a direct
fashion, we have attempted to measure
the stellar velocity dispersion in the central spectrum
by using the \ion{Ca}{2} K absorption line, which falls in a relatively
clean part of the spectrum. This was done by convolving with the
instrumental profile the average of three stars (two G0 IV and one F8 IV 
stars that fit the integrated stellar spectrum well) from the KPNO coude 
feed spectral atlas \citep{lei96}, then finding the Gaussian that would 
bring this profile up to the observed \ion{Ca}{2} K profile.  The resultant 
sigma was found to be 270 $\pm$ 50 km/s, corresponding to a massive
elliptical.

While the \ion{Ca}{2} K absorption line is not the optimal line to use in 
determinations of velocity dispersion, it is the only one available to us.  
Interstellar contamination is  unlikely to be significant in a galaxy that is 
dominated by relatively old stars. In addition, we 
get a good fit to the strength of the \ion{Ca}{2} K line from the spectral 
synthesis model, which leaves little room for additional ISM absorption.
The \ion{Ca}{2} H and K lines are, however, intrinsically broad, so they are less
sensitive indicators of velocity dispersion than are narrower lines.
\citet{kor82} found that inclusion of these lines in the determination of velocity
dispersions by the Fourier quotient technique resulted in a systematic
overestimate of the velocity dispersion by about 20\%, possibly due to
the effect of slight mismatches between the template and galaxy spectrum,
together with the steep continuum gradient in the region of the H and K 
lines. We believe that our direct convolution of template spectra (after flattening the 
nearby continuum), while lacking the precision of the Fourier quotient 
technique (as reflected in our error estimate), is not likely to introduce
significant systematic errors.

We can use  the present-day velocity dispersion---black hole mass relation 
\citep[\eg][]{tre02} to make an estimate of the 
mass of the 
black hole in 3C\,280, assuming that this relationship does not evolve with 
cosmic time. We have made a rough aperture correction using the formula
of \citet{jor95}
to the $r_e/8$ radius used by Tremaine et al., which raises our estimate of
sigma to 290 kms$^{-1}$, corresponding to a black hole mass of $\sim 6\times
10^{8} M_{\odot}$, typical for a radio galaxy in the local Universe.

We obtained another estimate of the black hole mass from the luminosity
of the galaxy, using the correlation of \citet{vdM99}.
We corrected the $H$-band magnitude of the galaxy to rest-frame $V$-band
using the elliptical galaxy model of \citet{fio97}.
Correcting for passive evolution of 0.9 magnitudes in $V$-band
using the same model, we estimate the black hole mass to be $\sim 7\times 
10^{8} M_{\odot}$, in good agreement with our estimate from the velocity 
dispersion of the galaxy.

\section{Discussion}

Our imaging and spectroscopic observations of 3C\,280
have allowed us to determine the probable sources of
the rest-frame optical and UV emission we observe from
most of the morphological components. 
We attempt to draw these data together
in combination with our high resolution radio mapping
to provide a coherent physical picture of the galaxy.

\subsection{Extended Emission-Line and UV Continuum Structure}

The rest-frame UV continuum structure mirrors the [\ion{O}{2}] image closely.
The most likely emission sources that provide
both line and continuum emission with similar morphologies
are (1) \ion{H}{2} regions with their associated young stars, and 
(2) line and thermal
emission from hot gas, produced either from external photoionization or
from shocks. 
We find no evidence for large populations of extremely young stars in our 
spectra that could produce \ion{H}{2} regions. On the other hand,
our observations indicate
that nebular continuum emission 
is a large contributor to all of the UV components, and the line ratios
discussed above tend to favor photoionization over shocks. In addition,
no large velocity gradients or line widths were found in any of the components,
which also indicates that direct shock heating of the gas is unlikely to
be important. 

In the eastern components, the [\ion{Ne}{3}]/[\ion{Ne}{5}] line ratios drop
off with distance from the nucleus, supporting photoionization as the
primary excitation mechanism in these components. 
The eastern components show morphological placement and kinematic 
signatures that make it likely that the gas has been lightly 
entrained in the outskirts of the radio lobe expansion,
yet is not being violently disturbed or heated by associated shocks. 

Some of the most intriguing features of the rest-frame UV 
image of 3C\,280 are the arc $d$
between components $b$ and $c$, and component $c$ itself.
The apparent edge brightening in $d$ and $c$ seen in the
HST F622W image likely results from
illumination effects involving photoionization and/or scattering.
The decomposition of the arc spectrum shows that
the continuum is almost entirely nebular thermal emission, and
the line emission is probably mostly due to photoionization from the quasar
nucleus, as there is no evidence for young stars in the arc.

To produce the unusual arc morphology, however,
the main possibilities are 
(1) that the arc is a tidal feature resulting from an
interaction between component $c$ (an associated galaxy, in this view)
and the central mass of 3C\,280,  that is then 
photoionized by the quasar; or (2) that the
arc morphology is formed by a shock front,
caused by the radio source or jet in some way. 
The closest analogy at low redshift to this arc that we are
aware of is the similarly smooth semicircular arc seen in [\ion{O}{3}]
emission in the $z=0.312$ quasar 3C\,249.1 \citep{sto83,sto87}. 
This arc also appears to be principally a gaseous feature, with no evidence
of associated stars.
This similarity, however, is not very illuminating, since the origin of 
this arc is
also uncertain. \citet{sto83} suggested a tidal origin, but it is also
true that the arc and a straight counter-feature are both roughly aligned 
along the radio axis.  If these features are tidal, their alignment 
with the radio axis would be solely a result of enhanced emission within the 
quasar opening angle. 
The arc ``echo'' in 3C\,280 (if real) might be difficult to account for in this
picture.  The placement of $c$ on the radio axis
would also have to be viewed as a coincidence, aside from its enhancement
due to scattering and photoionization.

Alternatively, the arc
(and perhaps the flattened component $c$) might be due
to emission at or behind
a shock front that originated somewhere near the radio axis.  Both
line and continuum emission could come from
the shock-heated gas, and are generally expected to occur
down-stream of the shock as the gas cools. Photoionization
of the material then could also occur after the shocked gas bubble had formed.
The arc ``echo'' might then be a natural second feature 
caused by whatever shock system created the arc in the first place.
However, while we have no way of easily discriminating shock
ionization from pure photoionization in the arc, 
the edge brightening and the low velocities relative to the 
nucleus lead us to believe that this gas is not currently being severely
affected by shock. Perhaps some of the relative velocity in the 
arc could be hidden by making it a transverse velocity, but the
velocities in the gas are certainly very quiescent.

An emission-line arc, though not nearly as
precise and circular as the one in 3C\,280,
is seen around the radio lobe of the low-$z$
radio galaxy PKS 2250-41 \citep{tad94}. The arc is associated
with blue continuum emission, and is likely to have
come from a shock caused by the impact of the radio jet
on a region of ambient material with higher-than-average density.
However, in that case, the arc clearly follows the lobe radio contours;
in 3C\,280, the arc is not associated with a radio lobe at all (unlike
the material on the east side, which does appear to be related to the
eastern radio lobe). 

If we had detected a stellar component in the arc $d$ with ages older
than the likely age of the radio source, we would have had fairly
compelling evidence for a tidal origin for this feature.  As it is,
we cannot come to any such firm conclusion: the arc could indeed be a
tidal feature, but it seems equally possible that it is some sort of
shock phenomenon associated with the radio jet.  The similarity to
the arc associated with 3C\,249.1 remains striking, and it may be that
a more detailed study of this object would prove illuminating for
our understanding of 3C\,280.

In $c$, while the [\ion{O}{3}] $\lambda\lambda4959$,5007/[\ion{O}{3} ]$\lambda$4363 line
ratio we observe is more consistent with photoionization
than with pure shock excitation, it is certainly consistent with
combined models, in which shocks ionize pre-shock gas.
A definite problem with a shock formation interpretation is the
presence of stars with ages of $\sim100$ Myr.  These are probably
too old to be
consistent with jet-induced star formation associated with the
current radio source, and it is difficult to believe that the
alignment would be preserved if they had been produced in a previous
radio outburst.
However, the gap in $c$ at the longer wavelengths aligns quite 
well with the jet; this is reminiscent of the jet-associated
gap in Cygnus A \citep{jac94,cab96}. 
If this rest-frame 0.8 $\mu$m morphology is associated
with the young stellar component, and we take the
gap as jet-associated, then it is difficult to explain this
gap without invoking jet-induced star formation. 

In the $z \sim 1$ radio galaxy 3C356, a discrete aligned component 
showing a 4000 \AA\ break falls directly on the jet axis, yet has
infrared-to-optical
colors that are too red to be consistent with a stellar population
young enough to be the result of jet-induced star formation from the 
current epoch of radio activity \citep{lac94}.
In this case, \citet{lac94} suggest that the stellar population
induced by the jet might have had an initial mass function (IMF)
heavily weighted towards the high mass end, resulting in a population
(after a few $\times 10^7$ years) dominated by red supergiants. 
This would allow for red colors
in less than the lifetime of the radio source. Such top-heavy IMFs
are observed in starburst galaxies, and have been suggested 
as a possible consequence of jet-induced star formation \citep{rees89,bit90}.
In our case, we see no obvious signatures of very young massive stars in 
our spectrum of $c$. 
In fact, a large contribution from red supergiants would hurt,
rather than help, our decomposition of its spectrum, and we would still
need to identify another blue component to allow us to match the observed
spectrum.  However, we cannot exclude the possibility that
by varying metallicity effects and using high mass stars with
a range of evolutionary states
we could produce an integrated spectrum close to
that we have identified as a $\sim100$-Myr-old population.

Morphologically,
we see some structures that must be related to the radio lobe and jet
expansion and interaction with the galaxy and ambient medium, yet
line diagnostics and kinematics give little evidence of shocks.
Possibly most of the aligned emission line gas in 3C\,280 has passed
through the shocks, and is now cooling and being photoionized. 
This is consistent with 3C\,280 playing some intermediate role
between the compact steep-spectrum objects that \citet{bes00b} say
are primarily shock-excited, and the extended, less aligned,
radio galaxies from the same sample which are dominated by photoionization.
This interpretation is also consistent with  3C\,280's intermediate
radio size (and therefore intermediate age of radio activity) in 
the sample of \citet{bes00a}.

\subsection{The Elliptical Galaxy}

Although much of the aligned light is in the rest-frame UV components,
the underlying old elliptical also appears to be extremely well aligned with
the radio axis.  The elliptical is well fit by a de Vaucouleurs profile both
in the high resolution, rest-frame 0.8 $\mu$m NICMOS image and at rest-frame
1 $\mu$m from a deep Keck image. 
Although we cannot rule out from the imaging alone the possibility of
a contribution from some extended aligned, red component with a similar 
spatial profile, spectrocopy indicates that the stellar
population should be fairly dominant at rest-frame $\lambda$ $>$ 4000\AA,
and we see no evidence for a non-stellar red component. 

The stellar population is clearly fairly
old; our spectroscopic fit gives a range of about 2--4 Gyr, assuming solar
metallicities. We also find that the galaxy has a high mass, with a stellar
velocity dispersion $\sigma=270$ km s$^{-1}$.
On any reasonable interpretation of the evidence, this is a massive,
old, relaxed galaxy, fully in place by $z = 1$, and it is difficult
to understand the quite precise alignment of this distribution
of old stars with the radio axis and with other morphological components
such as the gap in object $c$ and the baseline of the arc $d$.
The easiest way out is simply to say that this is a single example, and
that the alignment is pure chance.
But, at low redshifts, \citet{roc00} find a similar alignment for the three
most powerful radio galaxies (3C\,330, 3C\,341, and 3C\,348) in their
sample of 16 3C galaxies.  This alignment is especially compelling for
3C\,348 (see also West 1994), which is the most powerful radio galaxy
at low redshift besides Cygnus A.  For Cyg A itself, the inner optical structure
has long been known to be aligned with the radio axis, although the
large-scale stellar distribution is not (its axis makes an angle of about
54\arcdeg\ with respect to the radio axis).  
At intermediate redshifts, \citet{lac99} found significant alignment of
a low-radio-luminosity sample of FR\,II radio galaxies at $z\sim0.8$ with
the radio axis; in many cases this alignment was clearly due to stellar
emission rather than to AGN-related emission.  At high redshifts,
\citet{zir03} find evidence for significant alignment of the rest-frame optical
host galaxies in their $H$-band NICMOS study of a sample of 3C $z\sim1$--2
radio galaxies, including 3C\,280. (Of this sample of 9 objects,
3C\,280 exhibited the closest alignment to the radio axis, so it
may indeed simply be an unusual example). 
However, careful studies of dust disks (often assumed to be perpendicular
to the rotation axis of a galaxy) and major axes of low-$z$ 3C host galaxies
using the HST 3C snapshot survey have shown no preferential alignment
or misalignment of the disk with respect to the jet axis in FR\,II sources
\citep{dek00}, especially if corrected for the three-dimensional orientation
of the jets \citep{sch02}.
Thus the evidence is mixed, although there is sufficient
cause to believe that the effect might be real to justify some speculation
on how it might have come about.

In order to produce an alignment between the large-scale stellar distribution,
there must be some mechanism that would coordinate the alignment of the rotation
axis of the central black hole with the formation of the spheroid.  The
relation between black-hole mass and the mass of the spheroid
\citep{fer00,geb00,tre02}
demonstrates that it is likely that the formation of these two components
is closely coupled, although it seems far easier to imagine
mechanisms for obtaining simply the mass correlation than ones that
give both the mass correlation and a preferential alignment between the
black-hole axis and the morphology of the spheroid.
Under the standard cold dark matter (CDM) scenario, elliptical galaxies are
formed from successive mergers of smaller units.  Nevertheless, they cannot be
{\it too} small without running afoul of the well-known color
(\ie\ metallicity)---luminosity relation for ellipticals
\citep{bow92,ell96}.  Under this picture,
they must form from a relatively small number of major mergers
\citep[\eg][]{kau98}.  It seems unlikely that, in such mergers,
the rotation axis of the final central black hole (if produced 
mainly from successive
mergers of black holes associated with the merging galaxies) will be
preferentially aligned with the final stellar distribution.  In fact,
one would expect that typical high-angular-momentum mergers might well
produce an anticorrelation.

This argument assumes that findings relevant to elliptical galaxies in
general can be applied without reservation to powerful radio galaxies at
high redshift.  However, such objects are extremely rare:  there are only
about a dozen radio galaxies in the currently observable Universe with the
radio power of 3C\,280 or greater at redshift less than 1.5.  Such rare
objects may have had a special history.  Furthermore, if material around
the parent population of radio galaxies is preferentially distributed along
the same axis as that of the stellar distribution, there will be a bias in
favor of detecting radio sources for which the radio axis lies along this
same direction \citep{eal92}.

If radio galaxies at
high redshifts are among the precursors of the brightest cluster galaxies (BCGs)
at low redshift \citep{wes94}, we can tap into the various arguments
that the formation of BCDs has indeed differed substantially from that of
other cluster galaxies \citep{san76,tre77,vdB83}.  In particular, \citet{wes94}
shows that the inner regions of BCGs are typically aligned with their
large scale environments, and he proposes that they have been formed quite
early from subgalactic clumps that have merged preferentially along the axes of
dark-matter filaments.  Such mergers, West argues, will tend to have
low angular momentum, and they will result in prolate stellar structures
in which gas will eventually settle into a disk at the center, rotating
with an axis aligned with that of the stellar distribution.
Under these circumstances, if the gas from this disk is being accreted onto
a central black hole at a reasonable rate, the axis of the black hole will
be brought into alignment with the axis of the prolate spheroid fairly
quickly ($\lesssim10^8$ years for typical parameters; \citealt{ree78,sch96,
nat98}).
Note that these arguments do not depend critically on the initial formation
scenario.  All we really need is (1) a stellar distribution in the form
of a prolate ellipsoid, (2) sufficient gas settling into a disk at the
center and forming or accreting onto a black hole, (3) enough time for the
torque of the stellar mass distribution to align the axis of the disk
with the axis of the galaxy, and (4) similarly, enough time for
the disk to align the rotation axis of the black hole with the 
disk axis. 

\section{Summary}

The key points of this paper, based on our analysis of new radio maps,
optical and IR imaging data, and spectroscopy of 3C\,280 are:

\begin{enumerate}
\item We confirm the close association of the extended emission regions
on the east side of 3C\,280 with the eastern radio structure.  These
are almost certainly a result of the interaction of the radio jet and
lobe material with ambient gas, even though the velocities and line widths
do not show obvious kinematic signatures of such an interaction.
\item The radio structure on the western side shows evidence of beamed
emission, both at the base of the jet and in the strongly peaked hot spots
associated with the western lobe.  This beaming is consistent with previous
evidence that places the western lobe nearer to us.
\item The spectrum of the central component and of all of the extended
emission components appear to be more consistent with photoionization by
the quasar nucleus than with shock ionization, although mixed models
cannot be excluded.
\item In four regions the spectrum can be decomposed into various
combinations of four basic components: emission lines and associated nebular
thermal continuum, the old stellar population comprising the elliptical
galaxy, a younger stellar population, and scattered quasar light.
In particular, component $c$ shows, in addition to strong emission lines and
nebular thermal continuum (and probably scattered quasar light), a stellar
population $\sim100$ Myr old.  This age is difficult to reconcile with
jet-induced star formation.
\item The light from the arc appears to be strongly dominated by gaseous
emission, with little or no light from stars and little, if any, scattered
quasar light.  We have not been able to resolve the physical origin of the
arc morphology---the most plausible alternatives are that it is either
a gaseous tidal tail or a shock front due to the passage of the radio jet.
\item The unresolved core ($a$) in the F622W image
noted by RS97 can be identified with a quasar spectral
component clearly present in the central spectrum; \ie\ we may have
a direct view of a highly obscured quasar nucleus, or (more likely) at least of
quasar light scattered from a region very close to the nucleus.
\item The spheroidal distribution of old stars in 3C\,280 is very well fit
by an $r^{1/4}$-law profile, with an effective radius of $\sim6$ kpc.  The
major axis of the galaxy appears to be
aligned with the radio axis to an accuracy better
than 5\arcdeg. If this type of close alignment is found in 
larger samples of objects, this result may affect our understanding 
of how powerful radio galaxies are formed. 
\end{enumerate}

\acknowledgments

We thank Don Osterbrock for his assistance with some of the line 
identifications, and we thank the anonymous referee for detailed
comments that helped us improve both the content and the presentation.  
The authors recognize the very significant
cultural role that the summit of Mauna Kea has within the indigenous
Hawaiian community and are grateful to have had the opportunity to
conduct observations from it.
Support for proposals GO-5401 and GO-6491 was provided by 
NASA through a grant from the Space Telescope Science Institute, which is 
operated by the Association of Universities for Research in Astronomy, Inc., 
under NASA contract NAS 5-26555.
Some of the data presented herein were obtained at the W. M. Keck Observatory,
which is operated as a scientific partnership among the California Institute
of Technology, the University of California, and the National Aeronautics and
Space Administration, and which was made possible by the financial support of
the W. M. Keck Foundation.
This work was partly carried out at the Jet Propulsion Laboratory,
California Institute of Technology, under contract with NASA.
This work was also supported in part by NASA LTSA grant NAG 5-10762,
and some of the ground-based observations were supported by NSF
grant AST 95-29078.


\newpage


\begin{deluxetable}{l c c c c c}
\tablenum{1}
\tablewidth{0pt}
\tablecaption{Imaging Observations of 3C\,280}
\tablehead{
\colhead{Telescope} & \colhead{Detector} & 
\colhead{Filter} & \colhead{Exposure} & \colhead {Pixel Scale } &
\colhead{Data Quality} 
}

\startdata
CFHT & Orb2048 & 7449/31 & \phn4 $\times$ 2700 s & 0\farcs087 & 0\farcs66 \\ 
Keck & InSb256 & $K^\prime$ & 27 $\times$ 96 s\phn\phn& 0\farcs15\phn & 0\farcs58\\
HST & WFPC2 & F622W & \phn8 $\times$ 1100 s& 0\farcs10\phn & 0\farcs14 \\
HST & WFPC2 & F814W & \phn2 $\times$ 1100 s& 0\farcs10\phn & 0\farcs15 \\
HST & NICMOS & F160W & 5 $\times$ 1026 s, 300 s, 514 s & 0\farcs075 & 0\farcs15 \\

\enddata
\label{obslog}
\end{deluxetable}


\begin{deluxetable}{l c c c c c c c c}
\tabletypesize{\scriptsize}
\scriptsize
\tablenum{2}
\tablewidth{0pt}
\tablecaption{Emission Line Fluxes: Central Components}
\tablehead{
\colhead{} & \multicolumn{4}{c}{Nucleus}& \multicolumn{4}{c}{W1} \\
\cline{2-5} \cline{6-9}
\colhead{Line} &
\colhead{$\lambda_{\rm obs}$(\AA)} &
\colhead{FWHM (\rm \AA) } &
\colhead{Flux\tablenotemark{a}} & \colhead{EW (\AA)} &
\colhead{$\lambda_{\rm obs}$(\AA)} &
\colhead{FWHM(\rm \AA) } &
\colhead{Flux\tablenotemark{a}} & \colhead{EW (\AA)}
}

\startdata

\ion{C}{2}] $\lambda2325$ & 4646.3 & 16.9 & \phn61.9 & \phn31.3 & 4644.4 & 15.2 & \phn39.8 & \phn30.8 \\
\ion{He}{2} $\lambda2385$ & 4762.4 & 13.1 & \phn\phn9.5 & \phn\phn4.8 & \nodata & \nodata & \phn\phn$<$4\phd\phs & \phn\phn$<$3.2\phs \\
\phm{}[\ion{Ne}{4}] $\lambda2424$ & 4841.0 & 14.1 & 122\phd\phn & \phn63.5 & 4838.5 & 11.9 & \phn87.0 & \phn63.2 \\
\phm{}[\ion{O}{2}] $\lambda2470$ & 4935.5 & 12.4 & \phn12.7 & \phn\phn6.5 & 4931.6 & \phn7.9 & \phn\phn7.8 & \phn\phn6.1 \\
\ion{He}{2} $\lambda2511$ & 5013.5 & 16.2 & \phn10.9 & \phn\phn5.6 & 5013.4 & 12.5 & \phn\phn9.7 & \phn\phn7.7 \\
\phm{}[\ion{Mg}{7}] $\lambda2629$ & 5250.8 & 14.7 & \phn11.1 & \phn\phn5.4 & 5245.5 & 15.4 & \phn10.1 & \phn\phn7.6 \\
\ion{He}{2} $\lambda2733$ & 5462.1 & 21.6 & \phn25.8 & \phn12.2 & 5457.5 & 11.4 & \phn12.0 & \phn\phn9.5 \\
\phm{}[\ion{Mg}{5}] $\lambda2783$ & 5560.6 & 15.2 & \phn20.9 & \phn\phn9.7 & 5555.1 & \phn7.0 & \phn\phn8.4 & \phn\phn6.3 \\
\ion{Mg}{2} $\lambda2796$\tablenotemark{b} & 5585.2 & 15.2 & 110\tablenotemark{c}\phd\phn & \phn51.4 & 5581.4 & 13.8 & \phn43.2 & \phn30.3 \\
\ion{Mg}{2} $\lambda2803$ & 5599.8 & 15.2 & .. & .. & 5596.1 & 13.8 & \phn34.9 & \phn24.5 \\
\phm{}[\ion{Fe}{4}] $\lambda2834$ & 5662.6 & 26.3 & \phn31.0 & \phn14.7 & 5658.1 & 33.9 & \phn21.0 & \phn15.5 \\
\ion{O}{3} $\lambda3133$ & 6260.7 & 18.7\tablenotemark{d} & \phn24.9 & \phn11.7 & 6254.0 & \phn8.8 & \phn\phn7.4 & \phn\phn5.8 \\
\ion{He}{2} $\lambda3203$ & 6398.4 & 15.9 & \phn28.2 & \phn12.8 & 6394.5 & 11.7 & \phn20.4 & \phn15.5 \\
\phm{}[\ion{Ne}{5}] $\lambda3346$ & 6683.7 & 17.9 & \phn68.6 & \phn31.3 & 6678.7 & 10.0 & \phn35.2 & \phn27.1 \\
\phm{}[\ion{Ne}{5}] $\lambda3426$ & 6843.9 & 17.2 & 168\phd\phn & \phn73.7 & 6838.5 & 10.3 & \phn90.0 & \phn65.5 \\
\phm{}[\ion{Fe}{7}] $\lambda3585$ & 7167.4 & 19.6 & \phn11.0 & \phn\phn5.1 & \nodata & \nodata & \nodata & \nodata \\
\phm{}[\ion{O}{2}] $\lambda\lambda3726$,3729 & 7444.8 & 20.0 & 594\phd\phn & 307\phd\phn & 7442.0 & 16.8 & 469.1 & 355.0 \\
\phm{}[\ion{Fe}{7}] $\lambda3758$ & 7510.0 & 26.9 & \phn21.1 & \phn11.7 & 7502\phd\phn & $\sim$17\phd\phn\phs & \phn\phn$\sim$7.2\phs & \phn\phn$\sim$5.7\tablenotemark{e}\phs \\
H10 $\lambda3798$ & \nodata & \nodata & \nodata & \nodata & 7580.3 & 10.8 & \phn\phn9.7 & \phn12.2\tablenotemark{e} \\
H9 $\lambda3835$ & \nodata & \nodata & \nodata & \nodata & 7655.6 & 11.1 & \phn15.2 & \phn16.6 \\
\phm{}[\ion{Ne}{3}] $\lambda3869$ & 7728.0 & 20.9 & 247\phd\phn & 122\phd\phn & 7722.9 & 12.9 & 186.3 & 173.7 \\
H8 $\lambda3889$ & 7767.3 & 23.2 & \phn42.3 & \phn20.8 & 7762.9 & 12.9 & \phn31.0\tablenotemark{f} & \phn26.6 \\
\phm{}[\ion{Ne}{3}] + H$\epsilon$  $\lambda3968$, $\lambda3972$ & 7926.6 & 21.6 & \phn94.9 & \phn44.2 & 7922.2 & 16.8 & \phn95.3 & 106.5 \\
\phm{}[\ion{S}{2}] $\lambda\lambda4069$,4076 & 8132.0 & 34.1 & \phn26.8 & \phn10.9 & 8125.6 & 21.2 & \phn17.1 & \phn14.9 \\
H$\delta$ $\lambda4101$ & 8192.8 & 22.8 & \phn65.0 & \phn26.6 & 8188.5 & 14.1 & \phn55.8 & \phn46.3 \\
H$\gamma$ $\lambda4340$ & 8669.5 & 24.8 & 142.0 & \phn57.9 & 8188.5 & 14.1 & \phn55.8 & \phn46.3 \\
\phm{}[\ion{O}{3}] $\lambda4363$ & 8715.2 & 28.4 & \phn53.2 & \phn21.7 & 8710.3 & 17.5 & \phn37.5 & \phn33.2\\

\\

\ion{Mg}{2} absorption $\lambda2796$ & 5476.4 & \phn4.3 & & \phn1.4 & &   & & \\
\ion{Mg}{2} absorption $\lambda2803$ & 5491.0 & \phn3.2 & & \phn1.3 & & & & \\
\ion{Ca}{2} absorption $\lambda3933$ & 7856.2 & 36.9\tablenotemark{g} & & 13.5 & & & &
\enddata\label{tabcen}
\tablenotetext{a} {Flux units are 10$^{-18}$ \ergflx.}
\tablenotetext{b} {Widths set equal in deblending of \ion{Mg}{2} $\lambda\lambda2796$,2803.}
\tablenotetext{c} {Sum of both \ion{Mg}{2} $\lambda\lambda2796$,2804.}
\tablenotetext{d} {Gaussian fit gives flux 37 (in these units); flux error is
systematic and larger than average.}
\tablenotetext{e} {Flux and EW variations $\sim2$--3 and $\sim3$--4 respectively.}
\tablenotetext{f} {Flux and EW variations $\sim5$ and $\sim4$--10 respectively.}
\tablenotetext{g} {Problems with sky lines; Gaussian fit gives FWHM 28.3, EW 9.91.}
\end{deluxetable}


\begin{deluxetable}{l c l c c c c }
\tabletypesize{\scriptsize}
\scriptsize
\tablenum{3}
\tablewidth{0pt}
\tablecaption{Emission Line Fluxes: Eastern Components}
\tablehead{
\colhead{Component} &
\colhead{Offset\tablenotemark{a}} &
\colhead{Line} &
\colhead{$\lambda_{\rm obs}$(\AA)} &
\colhead{FWHM (\rm \AA) } &
\colhead{Flux\tablenotemark{b}} & \colhead{EW (\AA)} 
}

\startdata

E1a & 2\farcs24 &\ion{C}{2}] $\lambda2325$ & 4644.6 & 21.2 & \phn19.2 & \phn\phn88.9\\
 &  & \ion{He}{2} $\lambda2385$ & 4761.8 & \phn9.2 & \phn\phn4.3 & \phn\phn17.6\\
 &  & \phm{}[\ion{Ne}{4}] $\lambda2424$ & 4838.9 & 11.5 & \phn19.4 & \phn119.3 \\
 &  & \ion{Mg}{2} $\lambda2796$\tablenotemark{c} & 5586.9 & \phn7.2 & \phn\phn5.3 & \phn\phn25.0 \\
 &  & \ion{Mg}{2} $\lambda2803$\tablenotemark{c} & 5596.0 & \phn7.2 & \phn\phn5.4 & \phn\phn25.0 \\
 &  & \ion{He}{2} $\lambda3203$ & 6395.8 & \phn9.7 & \phn\phn5.2 & \phn\phn24.6 \\
 &  & \phm{}[\ion{Ne}{5}] $\lambda3346$ & 6684.6 & 18.4 & \phn\phn6.1 & \phn\phn25.7 \\
 &  & \phm{}[\ion{Ne}{5}] $\lambda3426$ & 6840.5 & 16.2 & \phn13.6 & \phn\phn56.7 \\
 &  & \phm{}[\ion{O}{2}] $\lambda\lambda3726$,3729 & 7443.6 & 15.1 & 125.0 & \phn473\phd\phn \\
 &  & H9 $\lambda3835$ & 7660.3 & 19.7 & \phn\phn4.8 & \phn\phn33.4 \\
 &  & \phm{}[\ion{Ne}{3}] $\lambda3869$ & 7725.4 & 14.3 & \phn47.8 & \phn225\phd\phn \\
 &  & H8 $\lambda3889$ & 7767.5 & 11.8 & \phn\phn8.6 & \phn\phn48.4 \\
 &  & \phm{}[\ion{Ne}{3}] + H$\epsilon$  $\lambda3968$, $\lambda3972$ & 7924.2 & 13.3 & \phn21.8  & \phn112.4 \\
 &  & H$\delta$ $\lambda4101$ & 8190.3 & 14.1 & \phn16.7 & \phn\phn76.9 \\
 &  & H$\gamma$ $\lambda4340$ & 8666.4 & 22.6 & \phn39.3 & \phn207.1 \\
E1b & 3\farcs10 &\ion{C}{2}] $\lambda2325$ & 4646.2 & 13.9 & \phn11.7 & \phn\phn91.4\\
 &  & \phm{}[\ion{Ne}{4}] $\lambda2424$ & 4841.7 & \phn9.6 & \phn\phn6.4 & \phn\phn46.7 \\
 &  & \ion{Mg}{2} $\lambda2796$\tablenotemark{c} & 5587.0 & \phn7.2 & \phn\phn7.3 & \phn\phn69.4 \\
 &  & \ion{Mg}{2} $\lambda2803$\tablenotemark{c} & 5600.3 & \phn7.2 & \phn\phn3.2 & \phn\phn28.1 \\
 &  & \phm{}[\ion{Ne}{5}] $\lambda3426$ & 6837\phd\phn & 19\phd\phn & \phn\phn4.5 & \phn\phn33\phd\phn \\
 &  & \phm{}[\ion{O}{2}] $\lambda\lambda3726$,3729 & 7446.1 & 15.8 & \phn\phn9.7 & \phn833\phd\phn \\
 &  & \phm{}[\ion{Ne}{3}] $\lambda3869$ & 7727.9 & 14.1 & \phn26.0 & \phn251.0 \\
 &  & H8 $\lambda3889$ & 7767.5 & 19.4 & \phn\phn7.7 & \phn\phn69.7 \\
 &  & \phm{}[\ion{Ne}{3}] + H$\epsilon$  $\lambda3968$, $\lambda3972$ & 7928.7 & 13.5 & \phn10.2  & \phn127\phd\phn \\
 &  & H$\delta$ $\lambda4101$ & 8196.1 & 13.2 & \phn\phn9.9 & \phn151\phd\phn \\
 &  & H$\gamma$ $\lambda4340$ & 8665.6 & 21.9 & \phn15.1 & \phn\phn73\tablenotemark{d}\phd\phn \\
E1c & 3\farcs96 &\ion{C}{2}] $\lambda2325$ & 4649.1 & 13.1 & \phn10.4 & \phn150\phd\phn\\
 &  & \phm{}[\ion{Ne}{4}] $\lambda2424$ & 4842.3 & 10.5 & \phn\phn4.7 & \phn\phn70.4 \\
 &  & \ion{Mg}{2} $\lambda2796$\tablenotemark{c} & 5588.0 & \phn7.7 & \phn\phn5.0 & \phn\phn75.9 \\
 &  & \ion{Mg}{2} $\lambda2803$\tablenotemark{c} & 5601.8 & \phn7.7 & \phn\phn4.1 & \phn\phn60.8 \\
 &  & \phm{}[\ion{O}{2}] $\lambda\lambda3726$,3729 & 7449.8 & 18.8 & \phn79.6 & 2206\phd\phn \\
 &  & \phm{}[\ion{Ne}{3}] $\lambda3869$ & 7733.3 & 17\phd\phn & \phn10.3  & \phn140\phd\phn \\
 &  & \phm{}[\ion{Ne}{3}] + H$\epsilon$  $\lambda3968$, $\lambda3972$ & 7932.6 & 11.4 & \phn\phn6.9  & \phn391\phd\phn \\
E2a & 5\farcs10 & \ion{C}{2}] $\lambda2325$ & 4652.9 & 11.4 & \phn13.4 & \phn172\phd\phn \\
 & & \ion{He}{2} $\lambda2385$ & 4765.7 & \phn7.8 & \phn\phn3.9 & \phn\phn49\phd\phn \\
& & \ion{Mg}{2} $\lambda2796$ & 5591.3 & 10.3 & \phn13.8 & \phn144\phd\phn\\
& & \ion{Mg}{2} $\lambda2803$ & 5606.9 & \phn9.9 & \phn\phn8.9 & \phn\phn95\phd\phn \\
& & \phm{}[\ion{Ne}{5}] $\lambda3426$ & 6849.8 & 10.2 & \phn\phn3.3 & \phn\phn29.5 \\
& & \phm{}[\ion{O}{2}] $\lambda\lambda3726$,3729 & 7456.0 & 14.5 & 135.3 & 1848\phd\phn\\
& & H9 $\lambda3835$\tablenotemark{e} & 7674.3 & 14.6 & \phn$<$3.8\phd & \phn\phn71\phd\phn \\ 
& & \phm{}[\ion{Ne}{3}] $\lambda3869$ \tablenotemark{e} & 7737.6 & 13.6 & \phn17.4 & \phn205\phd\phn\\
& & H8 $\lambda3889$ \tablenotemark{e} & 7780.5 & 14.6 & \phn$<$3.6\phd & \phn\phn55\phd\phn\\
& & \phm{}[\ion{Ne}{3}] + H$\epsilon$  $\lambda3968$, $\lambda3972$ & 7938.2 & 19.0 & \phn12.2  & \phn192\phd\phn\\
& & H$\delta$ $\lambda4101$ & 8202.2 & 11.7 & \phn\phn6.8  & \phn\phn45\phd\phn \\
& & H$\gamma$ $\lambda4340$ & 8680.3 & 12.0  & \phn11.3 & \phn102\phd\phn\\
E2b & 6\farcs11 & \ion{C}{2}] $\lambda2325$ & 4654.3 & 18\phd\phn & \phn\phn7.8 & \phn\phn86\phd\phn\\
& & \phm{}[\ion{Ne}{4}] $\lambda2424$ & 4841.4 & 18\phd\phn & \phn\phn6.1 & \phn\phn78\phd\phn\\
& & \ion{Mg}{2} $\lambda2796$\tablenotemark{c} & 5591.1 & \phn5.4 & \phn\phn2.9 & \phn\phn22\phd\phn\\
& & \ion{Mg}{2} $\lambda2803$\tablenotemark{c} & 5606.4 & \phn5.4 & \phn\phn2.8 & \phn\phn29\phd\phn \\
& & \phm{}[\ion{Ne}{5}] $\lambda3426$ & 6847.5 & \phn9.2 & \phn\phn4.3 & \phn\phn83\phd\phn \\
& & \phm{}[\ion{O}{2}] $\lambda\lambda3726$,3729 & 7455.5 & 14.3 & \phn53.6 & 1400\phd\phn\\
& & \phm{}[\ion{Ne}{3}] $\lambda3869$ & 7735.9 & 19.9 & \phn\phn8.0 & \phn137\phd\phn\\

\enddata\label{tabeast}
\tablenotetext{a} {Offset to the east (along slit) of center of aperture from the nucleus. Each aperture is 0\farcs86 wide.} 
\tablenotetext{b} {Flux units are 10$^{-18}$ \ergflx.}
\tablenotetext{c} {Widths set equal in deblending of \ion{Mg}{2} $\lambda\lambda2796$,2803.}
\tablenotetext{d} {H$\gamma$ region very noisy in E1B, 
poor continuum determination. }
\tablenotetext{e}{In E2A, H8 and H9 are detected, but because the continuum
is uncertain, they are fit together with [\ion{Ne}{3}] $\lambda3869$,
yielding upper limits to the fluxes. }
\end{deluxetable}


\begin{deluxetable}{l c c c c c}
\tablenum{4}
\tablewidth{0pt}
\tablecaption{Near-Infrared Emission Line Fluxes}
\tablehead{
\colhead{Component} & \colhead{Line} & \colhead{$\lambda_{\rm obs}$(\AA)}
& \colhead{Flux\tablenotemark{a}} & \colhead{FWHM(\AA)}  
}

\startdata
Central & [O\,III]$\lambda$5007 & \phn9989.6 & 4.46 & 20.3 \\
Central & [O\,III]$\lambda$4959 & \phn9889.1 & 1.39 & 26.8 \\
Central & H$\beta$ & \phn9709.8 & 1.47 & 37.3 \\
Central & H$\alpha$ & 13105\phd\phn & 1.96 & 32\phd\phn \\
W & [O\,III]$\lambda$5007 & \phn9990.5 & 2.69 & 18.9 \\
W & H$\beta$ & \phn9715.4 & 1.29 & 38.6 \\
E1 & [O\,III]$\lambda$5007 & \phn9992.0 & 2.18 & 26.8 \\
E2 & [O\,III]$\lambda$5007 & 10004.5 & 0.63 & \phn6.0 \\

\enddata
\label{irlinetab}
\tablenotetext{a} {Flux units are $10^{-15}$ erg cm$^{-2}$ s$^{-1}$.}
\end{deluxetable}

\end{document}